\documentclass[a4paper,onecolumn,11pt,accepted=2022-03-04]{quantumarticle}
\pdfoutput=1
\usepackage[utf8]{inputenc}
\usepackage[T1]{fontenc}
\usepackage{amsmath}
\usepackage{hyperref}
\usepackage{braket}
\usepackage{graphicx}
\usepackage{amssymb}
\usepackage[sorting=none]{biblatex}
\usepackage{tikz}
\usepackage{lipsum}

\begin{document}

\title{A Logarithmic Bayesian Approach to Quantum Error Detection}

\author{Ian Convy}
\email{ian\_convy@berkeley.edu}
\affiliation{Department of Chemistry, University of California, Berkeley, CA 94720, USA}
\affiliation{Berkeley Quantum Information and Computation Center, University of California, Berkeley, CA 94720, USA}
\orcid{0000-0003-1818-2677}
\author{K. Birgitta Whaley}
\orcid{0000-0002-7164-4757}
\affiliation{Department of Chemistry, University of California, Berkeley, CA 94720, USA}
\affiliation{Berkeley Quantum Information and Computation Center, University of California, Berkeley, CA 94720, USA}

\maketitle

\begin{abstract}
We consider the problem of continuous quantum error correction from a Bayesian perspective, proposing a pair of digital filters using logarithmic probabilities that are able to achieve near-optimal performance on a three-qubit bit-flip code, while still being reasonable to implement on low-latency hardware. These practical filters are approximations of an optimal filter that we derive explicitly for finite time steps, in contrast with previous work that has relied on stochastic differential equations such as the Wonham filter. By utilizing logarithmic probabilities, we are able to eliminate the need for explicit normalization and can reduce the Gaussian noise distribution to a simple quadratic expression. The state transitions induced by the bit-flip errors are modeled using a Markov chain, which for log-probabilties must be evaluated using a LogSumExp function. We develop the two versions of our filter by constraining this LogSumExp to have either one or two inputs, which favors either simplicity or accuracy, respectively. Using simulated data, we demonstrate that the single-term and two-term filters are able to significantly outperform both a double threshold scheme and a linearized version of the Wonham filter in tests of error detection under a wide variety of error rates and time steps.
\end{abstract}

\section{Introduction}

One of the major obstacles to large-scale quantum computation is the high frequency of qubit errors. Small interactions with the environment or imperfections in gate implementation can perturb the underlying quantum state throughout a computation and ultimately render the output useless. These issues only magnify as the size and complexity of the quantum computer increases, jeopardizing any attempt to demonstrate quantum advantage in key tasks such as the factorization of large integers~\cite{Shor_1997}. The obvious challenge posed by quantum errors has given rise to the field of \textit{quantum error correction} \cite{Lidar_Brun_2013} and spurred the development of numerous techniques to identify and correct errors when they occur. These schemes, referred to as error correction \textit{codes}, typically operate by encoding the quantum state into a larger Hilbert space and then monitoring the location of the state within this space using a specific set of observables.

The present work is focused on continuous or ``always-on'' error correction, where the state of the system is continuously monitored through a noisy signal channel in order to diagnose errors and to make corrections if necessary. Since the signal is noisy, we cannot know with certainty whether an error has occurred, and must instead make subjective judgements about the condition of the system based on past measurements and any other information we might have. In order to quantify the uncertainty in this task it is natural to adopt a Bayesian framework, which allows for new information to be easily combined with prior knowledge of the system using Bayes' theorem.

Assuming that the errors experienced by a quantum computer are uncorrelated, we can model the system as a Markov chain whose state is imperfectly observed by a measurement apparatus. Optimal treatments of this problem originate with work done by Wonham and his derivation of what is today known as the \textit{Wonham filter}, which describes the evolution of the Markov state probabilities conditioned on a particular set of noisy observations \cite{Wonham_1964}. Despite its optimality, this filter has not found widespread use for error correction, in part because it assumes that the value of the signal is characterized as a continuous function of time, while in practice the signal is typically only observed as a discrete set of measurement samples. As a non-linear stochastic differential equation, the Wonham filter can also be challenging to evaluate numerically, even if the signal function is known. Due to these limitations, most continuous error correction schemes avoid using Bayesian theory and instead rely on various thresholding procedures which are easier to implement but known to provide suboptimal performance \cite{Atalaya_Bahrami_Pryadko_Korotkov_2017}\cite{Mohseninia_Yang_Siddiqi_Jordan_Dressel_2020}\cite{Atalaya_Zhang_Niu_Babakhani_Chan_Epstein_Whaley_2021}.

The purpose of this work is to introduce a discrete-time filter for continuous quantum error correction that significantly outperforms common thresholding schemes, while still being practical to implement on real hardware. We construct our filter using logarithmic probabilities, which are numerically stable, easy to normalize, and allow for straightforward evaluation of Gaussian noise distributions using only arithmetic operations. The filter has two different forms based on how it updates the posterior, with one method emphasizing accuracy while the other emphasizes computational simplicity. Both versions of the filter are able to achieve near optimal performance under a wide variety of error rates and time steps.

The remainder of the paper has the following structure. Sec.~\ref{sec:quantum_error_correction} reviews the three-qubit code that we will use as a simple error correction primitive, and describes the principles of continuous measurement. Sec.~\ref{sec:optimal} derives an optimal Bayesian filter for identifying bit-flip errors, which cannot be practically implemented in its exact form. Sec.~\ref{sec:obstacles} discusses prior work in the field of continuous quantum error correction and outlines several obstacles to implementing a Bayesian filter. Sec.~\ref{sec:practical_filter} proposes an approximate version of the optimal filter which overcomes those obstacles, and provides two different implementations which are both experimentally feasible using today's technology. Sec.~\ref{sec:numerical} tests our filter on simulated data and compares the performances of the two implementations against other schemes found in the literature. Sec.~\ref{sec:discussion} summarizes our findings and identifies areas for future work. An \hyperref[sec:appendix]{Appendix} is also provided, which contains technical details that supplement the body of the paper.

\section{Quantum Error Correction}\label{sec:quantum_error_correction}

\subsection{Three-qubit coding scheme}\label{sec:error_correction}

To insulate a quantum system from errors, we must add some level of redundancy to its Hilbert space. This is achieved by assembling a set of physical qubits which is larger than necessary to perform the desired task, with the idea that these extra degrees of freedom will be used to implement an error correction scheme. The computation is then understood in terms of its effect on the \textit{coding subspace} of the system, which describes the portion of the expanded Hilbert space where the intended quantum evolution will take place.

For simplicity we focus on a three-qubit repetition code, which is designed to protect the state of a single qubit from the effects of bit-flip errors. This encoded qubit is the true object of interest, so the computational task is conceptualized in terms of single-qubit operations which then need to be mapped to the full three-qubit system. We can perform this mapping by considering the so-called \textit{logical qubit}, defined as
\begin{equation}\label{eq:logical_qubit}
    \ket{0_L} \equiv \ket{000} \text{ and }
    \ket{1_L} \equiv \ket{111},
\end{equation}
which occupies the two-dimensional subspace spanned by $\ket{000}$ and $\ket{111}$ that we designate as the coding subspace. 

The logical states in Eq.~\eqref{eq:logical_qubit} are used to encode the desired behavior of the system in the absence of errors, while the other parts of the Hilbert space serve to catch any bit-flips that occur. For example, a bit-flip error on the first qubit will transform $\ket{000}$ into $\ket{100}$ and $\ket{111}$ into $\ket{011}$, so the system will be shifted into a different two-dimensional subspace. As long as we can detect this change in the subspace, it will be possible for us to track and then correct the errors that have occurred on the system.

To perform this detection, we require a set of observables that uniquely identify the coding subspace and each of the three subspaces generated from bit-flips on the first, second, and third qubits. This is achieved using the \textit{parity operators} $Z_1Z_2$ and $Z_2Z_3$, whose eigenstates and eigenvalues are specified by
\begin{equation}\label{eq:syndrome_action}
    (Z_1Z_2\ket{\psi}, Z_2Z_3\ket{\psi}) =
    \begin{cases}
        (\ket{\psi}, \ket{\psi}) & \ket{\psi} = a\ket{000} + b\ket{111}
        \\
        (-\ket{\psi}, \ket{\psi}) & \ket{\psi} = a\ket{100} + b\ket{011}
        \\
        (\ket{\psi}, -\ket{\psi}) & \ket{\psi} = a\ket{001} + b\ket{110}
        \\
        (-\ket{\psi}, -\ket{\psi}) & \ket{\psi} = a\ket{010} + b\ket{101},
    \end{cases}
\end{equation}
for any complex numbers $a$ and $b$. Eq.~\eqref{eq:syndrome_action} shows that the combined action of the parity operators splits the three-qubit Hilbert space into four distinct 2-D subspaces, each characterized by a different pair of eigenvalues in $\{-1, 1\}$ which are referred to as \textit{syndromes}. The $(1, 1)$ syndrome identifies the coding subspace, while the other three combinations correspond to states which result from a bit-flip on one of the qubits. We refer to the latter as the \textit{error subspaces}, and note that the syndrome value identifies on which qubit the error has occurred.

The observables $Z_1Z_2$ and $Z_2Z_3$ offer us a simple method for bit-flip error detection, assuming that they can in fact be measured. After initializing the system into the coding subspace, we simply perform parity measurements of our system as needed and monitor the response of the system. If our results indicate that the system has moved out of the coding subspace, then we can record this deviation and apply a bit-flip correction based on which error subspace the system has been moved to. For example, if one error occurs on the first qubit then we have
\begin{equation}\label{eq:one_error}
    a\ket{000} + b\ket{111} \xrightarrow[\text{qubit 1}]{\text{error}} a\ket{100} + b\ket{011} \xrightarrow[\text{qubit 1}]{\text{correction}} a\ket{000} + b\ket{111},
\end{equation}
and the system has been returned to its proper state. Each time a bit-flip is diagnosed to have occurred on a given qubit, we can simply apply another bit-flip to that same qubit and the error will be undone.

It is important to note that if \textit{two} bit-flip errors occur simultaneously (or rapidly enough to appear indistinguishable) then the above detection scheme will mistakenly assume that only one error has occurred and perform the wrong correction. For example, if errors occur on the second and third qubits then
\begin{equation}\label{eq:two_errors}
    a\ket{000} + b\ket{111} \xrightarrow[\text{qubits 2,3}]{\text{errors}} a\ket{011} + b\ket{100} \xrightarrow[\text{qubit 1}]{\text{correction}} a\ket{111} + b\ket{000},
\end{equation}
and the system will experience a \textit{logical error} since the $a$ and $b$ coefficients have been exchanged. This type of misdiagnosis occurs because the errors in Eq.~\eqref{eq:one_error} and Eq.~\eqref{eq:two_errors} both move the system into the same parity subspace and thus return the same syndrome values, so we must respond with the same correction in both cases. Given that single errors are more common then double errors under realistic conditions, the most reasonable response is to always assume that one bit-flip has occurred whenever the system enters an error subspace. This compromise is an inescapable consequence of using a short code, though we can reduce its impact by detecting errors more quickly and thus better resolving the difference between one bit-flip and two sequential bit-flips.

Throughout this paper we will focus our attention on \textit{state tracking}, where the goal is to correctly identify the net number of bit-flips that have acted on the system at a given point in time. For simplicity, we will assume that the system is initialized into $\ket{000}$, and only evolves due to random bit-flip errors. That said, our results are not restricted to this quantum memory regime, as Eq.~\eqref{eq:syndrome_action} shows that the parity operators act uniformly on states within a given coding or error subspace, meaning that the measurement signal will be identical regardless of whether the system is in a basis state or an arbitrary superposition. Even if a Hamiltonian is applied to the system, as would occur during a process like quantum annealing \cite{Morita_Nishimori_2008}, the measurement signal will not be affected so long as the Hamiltonian does not mix together the four parity subspaces. As a result, any filter designed to handle quantum memory can be seamlessly applied to more complicated processes by just changing the metric used to evaluate its performance. 

\subsection{Error correction with continuous measurements}\label{sec:measurement}

The three-qubit code described in Sec.~\ref{sec:error_correction} is agnostic to the actual method used to probe $Z_1Z_2$ and $Z_2Z_3$, since the separation of the Hilbert space into coding and error subspaces is a property of the observables themselves rather than of any particular measurement scheme. Most work on quantum errror detection has focused on \textit{discrete} error correction, which utilizes projective measurements to resolve the exact syndrome of the state at periodic intervals \cite{Lidar_Brun_2013}. However, recent advances in superconducting qubit architecture has led to increased interest in \textit{continous} error correction, where the parity operators are monitored at all times and a noisy readout of the underlying syndrome values is generated \cite{Jacobs_Steck_2006}\cite{Korotkov_2016}. Under these noisy conditions, a filter will be needed in order to extract relevant information from the signal.

In an idealized setting, we can formulate a simple mathematical description of the continuous measurement process using Gaussian POVMs. At time $t$, the probability of observing signal readout $\alpha(t)$ from a parity operator with syndrome $S(t)$ is
\begin{equation}\label{eq:povm}
    \text{Prob}(\alpha(t)) \propto \exp\left[-\frac{(S(t) - \alpha(t))^2}{2\sigma^2}\right],
\end{equation}
which implies that the signal values are distributed as $\mathcal{N}(S(t), \sigma^2)$. The syndrome value $S(t) \in \{-1, +1\}$ is determined by the state of the system at time $t$, and will therefore change as errors occur. We assume that the variance $\sigma^2$ of the noisy signal is identical across all states and is thus independent of time. 

The weak measurement described by Eq.~\eqref{eq:povm} must occur over some finite time $\Delta t$, whose duration will determine the variance of the signal through the relation $\sigma^2 = \frac{k}{\Delta t}$ for some positive $k$ (the strength of the measurement is thus proportional to $\frac{1}{k}$). A longer measurement $\tilde{M}$ of time $T$ can be constructed by averaging over a set of these weaker measurements
\begin{equation}\label{eq:signal_avg}
    \Tilde{M}(t, T, \Delta t) = \left(\frac{T}{\Delta t}\right)^{-1}\sum^{\frac{T}{\Delta t}}_{n = 1}\alpha(t + n\Delta t) = \frac{1}{T}\sum^{\frac{T}{\Delta t}}_{n = 1}[S(t + n\Delta t) + \epsilon]\Delta t,
\end{equation}
where the signal at time increment $t + n\Delta t$ is the sum of the syndrome $S(t + n\Delta t)$ and a Gaussian noise term $\epsilon \sim \mathcal{N}(0, \frac{k}{\Delta t})$. The value of $t$ represents the time at which the averaging process starts, while the integer $n$ indexes the set of sequential weak measurements being averaged over, each of duration $\Delta t$. Since the average is taken over a time interval of length $T$, we can perform $\frac{T}{\Delta t}$ weak measurements within the interval. 

Given that the Gaussian noise $\epsilon$ in Eq.~\eqref{eq:signal_avg} is additive, its average can be calculated separately from the syndrome average as
\begin{equation}
    \xi(T) \equiv \frac{\Delta t}{T}\sum^{\frac{T}{\Delta t}}_{n = 1}\epsilon  = \frac{\Delta t}{T}\tilde{\epsilon},
\end{equation}
where $\tilde{\epsilon} \sim \mathcal{N}(0, \frac{kT}{(\Delta t)^2})$ due to the additive property of the variance and thus $\xi(T) \sim \mathcal{N}(0, \frac{k}{T})$ after scaling by $\frac{\Delta t}{T}$. The value of $\tilde{M}$ is therefore equal to the averaged syndrome plus a new noise term $\xi(T)$, whose variance is inversely proportional to $T$:
\begin{equation}\label{eq:discrete_avg}
    \Tilde{M}(t, T, \Delta t) = \frac{1}{T}\sum^{\frac{T}{\Delta t}}_{n = 1}[S(t + n\Delta t)\Delta t] + \xi(T).
\end{equation}
From Eq.~\eqref{eq:discrete_avg}, we define the \textit{continuous} measurement $M$ as the limit of $\tilde{M}$ for $\Delta t \rightarrow 0$, which turns the discrete sum into an integral,
\begin{equation}\label{eq:cont_avg}
    M(t, T) = \frac{1}{T}\int^{T}_0 S(t + \tau)d\tau + \xi(T),
\end{equation}
that is distributed as a Gaussian with variance $\frac{k}{T}$ and a mean value given by the integrated average of the syndrome over the interval.

\begin{figure}
    \centering
    \includegraphics[width=\textwidth]{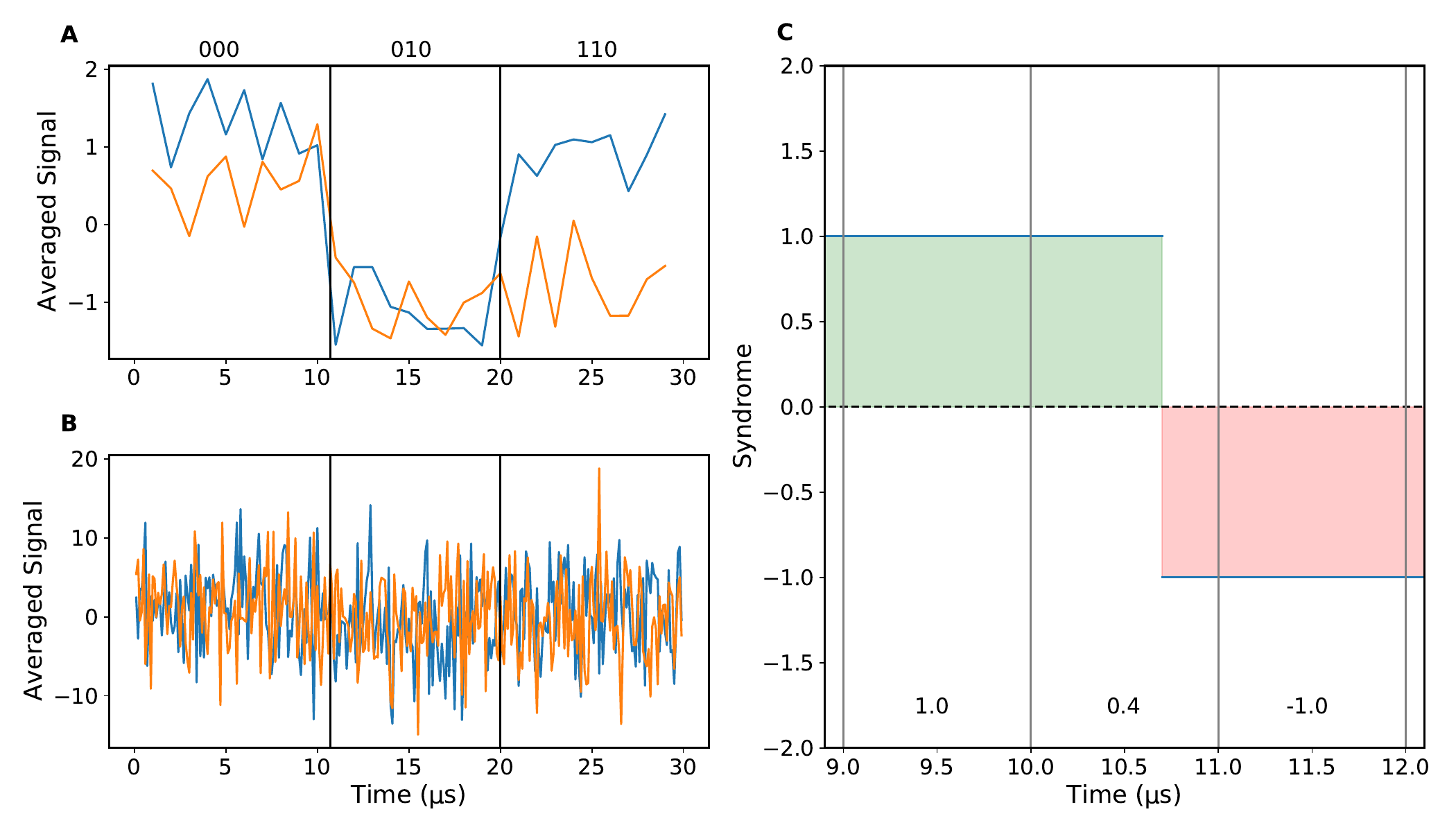}    \caption{Plots A and B show examples of 30 $\mu$s measurement sequences for $T = 1\ \mu\text{s}$ and $T = 0.1\ \mu\text{s}$ respectively, with $k = 0.5\ \mu\text{s}$ for both plots. Each point represents the unitless value of $M(t, T)$ taken at the time shown on the x-axis for one of the two parity operators, with the blue lines tracking the noisy measurement of $Z_1Z_2$ and the orange lines tracking $Z_2Z_3$. The underlying state evolution in both plots is identical, with the vertical black lines indicating the times at which bit-flip errors occurred (10.7 $\mu$s and 20 $\mu$s). The state of the system before and after each error is given by the three-bit number above plot A. Plot C shows the true syndrome value from $Z_1Z_2$ without noise, zoomed in on times near the first bit-flip. The vertical gray lines indicate the integration intervals used to generate the averaged syndromes found in Eq.~\eqref{eq:cont_avg} when $T = 1\ \mu\text{s}$, with the numerical values of these averages given at the bottom of the plot. The shaded regions indicate the ``area under the curve'' contribution at a given time, with green indicating a positive contribution and red a negative contribution. The bit-flip error occurs within the second interval, causing its syndrome average to deviate from the expected values of $\pm 1$ and instead lie somewhere in between.}
    \label{fig:syndromes}
\end{figure}

In Figure \ref{fig:syndromes} we provide examples of measurement sequences that could be observed as a three qubit system evolves under the influence of random bit-flip errors for 30 $\mu$s. For a system starting in $\ket{000}$, the measurements $M(t, T)$ for each syndrome will be centered at +1 and oscillate randomly due to the additive noise term. Comparing Figures \hyperref[fig:syndromes]{1A} and \hyperref[fig:syndromes]{1B}, it is clear that increasing the integration time $T$ from $0.1\ \mu$s (panel B) to 1 $\mu$s (panel A) dramatically reduces the fluctuations of the signal, making it much easier to see visually the effects of the bit-flip errors on the measurements. Perhaps counterintuitively, this reduction in the variance does not actually make the task of identifying a syndrome value from its noisy measurements any easier, since the standard error (SE) of mean value estimation scales with both the noise strength $\frac{k}{T}$ \textit{and} the number of samples to be averaged. Given that the number of measurements $n$ in a time interval of length $\beta$ is inversely proportional to $T$, i.e., $n = \frac{\beta}{T}$, we have
\begin{equation}\label{eq:standard_error}
    \text{SE}^2 = \frac{\sigma^2}{n} = \frac{k}{T}\frac{T}{\beta} = \frac{k}{\beta}
\end{equation}
and thus the uncertainty in our estimate of the syndrome will not depend on $T$.

This is not to say, however, that the length of the integration period has no effect on the state tracking problem. Indeed, Eq.~\eqref{eq:standard_error} simply states that there is a proportionate trade-off between the informativeness of a measurement ($\sigma^2=\frac{k}{T}$) and the number of such measurements ($n=\frac{\beta}{T}$) when the syndrome value is \textit{fixed}. When we introduce bit-flip errors into the system and thus consider situations where the syndrome changes with time, the analysis is significantly more complicated. Figure~\hyperref[fig:syndromes]{1C} shows that when an error occurs within an integration period, a portion of the syndrome average will consist of +1 contributions and the other portion will consist of -1 contributions. The resulting syndrome mean can therefore take on any value in the interval $[-1, +1]$, which suggests that the measurement will be inherently less informative than a measurement that occurs when the state of the system is constant. As the value of $T$ grows larger, more measurements will fall into this category. The incorporation of these intermediate syndrome means into an optimal state tracking model is explored in Sec.~\ref{sec:optimal}.

\subsection{Bayesian treatment of measurement}

Determining the state of a quantum system from noisy signals is not a new problem, and the field of quantum mechanics has long grappled with how to properly describe the mechanism and effects of measurement \cite{Wiseman_Milburn_2009}. One of the key challenges is knowing how to properly update the density matrix $\rho_s$ of a system to reflect the fact that we have performed a measurement on it. A common procedure involves coupling the system to a detector and then tracing out the detector's degrees of freedom, 
\begin{equation}\label{eq:trace_measurement}
    \rho_s = \text{Tr}_d(\rho_{sd})
\end{equation}
where $\rho_{sd}$ is the joint density matrix of the system and detector.
By its very nature this approach cannot offer any information about the state of the system after a \textit{specific} measurement, since we are averaging over all of the possible detector configurations without reference to the outcome that really occurred. Instead, what Eq.~\eqref{eq:trace_measurement} describes is the behavior of an \textit{ensemble} of systems after they are all measured, since the frequency of each measurement outcome can be specified without knowing the result for any given system. This approach is therefore insufficient for an error correction scheme, as the very purpose of error correction is to monitor a specific system and respond to the errors that actually occur, rather than to the average over all possible errors.

The issue of specific outcomes versus ensemble behavior is reminiscent of the disagreement between Bayesian and frequentist statistics \cite{Barnett_1999}, and it is therefore unsurprising that a Bayesian formalism for quantum measurement was developed. This formalism was pioneered by Korotkov, who proposed an update rule for the diagonal elements of the density matrix after a measurement $M$ using Bayes' theorem that has the form
\begin{equation}\label{eq:quantum_bayes_rule}
    \hat{\rho}_{ii} = \frac{P(M|i)\rho_{ii}}{\sum_j P(M|j)\rho_{jj}},
\end{equation}
where $\rho_{ii}$ is the original diagonal element acting as the prior and $\hat{\rho}_{ii}$ is the updated (posterior) element \cite{Korotkov_1999}. The term $P(M|i)$ describes the probability of observing measurement result $M$ given that the system is in state $i$. 

\section{An Optimal Bayesian Filter}\label{sec:optimal}

For our purposes, the significance of Korotkov's measurement formalism from Eq.~\eqref{eq:quantum_bayes_rule} lies in its treatment of the density matrix elements as classical Bayesian probability distributions, which allows us to leverage the existing toolkit of Bayesian statistics \cite{De_Finetti_Machi_Smith_De_Finetti_2017}\cite{Sivia_Skilling_2006} to develop an error correction algorithm. Fundamentally, the goal of this algorithm is to determine, using all available information, the \textit{most probable} state of the system at each time step in order to detect errors. As in Eq.~\eqref{eq:quantum_bayes_rule}, we seek to derive a posterior probability distribution for the state of our system after each measurement, using knowledge of the underlying bit-flip dynamics and of the Gaussian noise corrupting the measurement signal.

\subsection{Recursive form of the posterior probability}

For notational simplicity, we introduce the vector-valued quantity 
\begin{equation}\label{eq:vec_measurement}
    \vec{M}_i \equiv 
    \begin{bmatrix}
        M^{(1)}(iT, T) \\ 
        M^{(2)}(iT, T)
    \end{bmatrix},
\end{equation}
where $M^{(1)}$ and $M^{(2)}$ are measurements of the parity operators $Z_1Z_2$ and $Z_2Z_3$ respectively and $i$ is a non-negative integer. The time argument $iT$ appears because the sets of (non-overlapping) sequential measurements will be spaced out in increments of $T$. To denote the state of the system after the $i$th measurement, i.e., at time $t = (i + 1)T$, we employ the compound index $0 \leq \ell_i \leq 7$. Using this notation, we write the posterior probability of the $\ell_i$th state as
\begin{equation}\label{eq:posterior_def}
    \hat{P}(\ell_i) \equiv P(\ell_i|\vec{M}_0...\vec{M}_i),
\end{equation}
where $P(C|AB)$ is generically the probability distribution of variable $C$ given knowledge of the state of variables $A$ and $B$. Using the chain rule of probability, Eq.~\eqref{eq:posterior_def} can be put into a recursive form as
\begin{equation}\label{eq:bayes_deriv}
\begin{split}
    \hat{P}(\ell_i) & = \frac{P(\vec{M}_i\ell_i|\vec{M_0}...\vec{M}_{i-1})}{P(\vec{M}_i|\vec{M}_0...\vec{M}_{i-1})}
    \\
    &= \frac{1}{P(\vec{M}_i|\vec{M}_0...\vec{M}_{i-1})}\sum^7_{\ell_{i-1} = 0} \hat{P}(\ell_{i-1})P(\vec{M}_i\ell_i|\ell_{i-1}\vec{M_0}...\vec{M}_{i-1})
    \\
    &= \frac{1}{P(\vec{M}_i|\vec{M}_0...\vec{M}_{i-1})}\sum^7_{\ell_{i-1} = 0} \hat{P}(\ell_{i-1})P(\vec{M}_i|\ell_i\ell_{i-1}\vec{M_0}...\vec{M}_{i-1})P(\ell_i|\ell_{i-1}\vec{M_0}...\vec{M}_{i-1})
    \\
    &= \frac{1}{P(\vec{M}_i|\vec{M}_0...\vec{M}_{i-1})}\sum^7_{\ell_{i-1} = 0} \hat{P}(\ell_{i-1})P(\vec{M}_i|\ell_i\ell_{i-1})P(\ell_i|\ell_{i-1}),
\end{split}
\end{equation}
where in the last line we explicitly assume that $\vec{M}_i$ and $\ell_i$ are conditionally independent of $\{\vec{M}_0,...,\vec{M}_{i-1}\}$ given knowledge of $\ell_{i-1}$. This is equivalent to assuming that the state transitions depend only on the prior state (Markovian assumption), and that the additive measurement noise is uncorrelated across time. Since $P(\vec{M}_i|\vec{M}_0...\vec{M}_{i-1})$ has no explicit $\ell_i$ dependence and is not otherwise coupled to any of the terms in the sum, it will be ignored in our subsequent analysis and treated as simply a normalization factor.

If we assume that the system begins in a known state, which is reasonable at the start of a quantum experiment, then Eq.~\eqref{eq:bayes_deriv} provides a recipe for iteratively computing the probabilities of future states in light of new measurement results. The two quantities that must be solved for as part of this procedure are 
\begin{enumerate}
    \item $P(\ell_i|\ell_{i-1})$, which is the probability of jumping from state $\ell_{i-1}$ to state ${\ell_i}$ in time $T$ due to bit-flip errors, and
    
    \item $P(\vec{M}_i|\ell_i\ell_{i-1})$, which is the probability of measuring values $\vec{M}_i$ from the parity operators when we have knowledge of the system states at the beginning and end of the integration period.
\end{enumerate}
The remainder of this section will be dedicated to deriving explicit functional forms for these two probability terms.

\subsection{Analysis of the transition probability}\label{sec:markov}

Our treatment of bit-flip errors will assume that they act on the system independently of one another and occur at a fixed rate $\mu$ that is known to us. In saying that the errors are independent, we mean that an error occurring in one time interval gives us no information about whether an error will occur in any other non-overlapping interval. The number $e_k$ of such errors acting on the $k$th qubit in a time interval of length $T$ will obey the following Poisson distribution
\begin{equation}\label{eq:poisson}
    P(e_k) = \frac{(\mu T)^{e_k}}{e_k!}e^{-\mu T},
\end{equation}
where we emphasize that the error rate is identical across all three qubits.

In order to use Eq.~\eqref{eq:poisson} to derive a functional form for $P(\ell_i|\ell_{i-1})$, we must specify the number of errors needed to connect state $\ell_{i-1}$ to state $\ell_i$. Taking for example a system that starts as $\ket{000}$ at step $i-1$ and ends up as $\ket{100}$ at step $i$, corresponding to $\ell_{i-1} = 0$ and $\ell_i = 4$, it is clear that the first qubit must have been flipped at some point in the interval between the two steps. However, this does not mean that only a single error occurred on the first qubit, or that none of the other qubits experienced any errors. Rather, it means that the first qubit experienced an \textit{odd} number of errors while the rest of the qubits experienced an \textit{even} number of errors (including no error at all), such that the net number of bit-flips works out to be one for the first qubit and zero for the others. We can therefore express $P(\ell_i|\ell_{i-1})$ as a sum over the Poisson probabilities of every possible error combination consistent with the net number of flips between $\ell_i$ and $\ell_{i - 1}$.

Rather than evaluate this sum directly, it is easier to look at each qubit separately and calculate the probability that it will experience either an even or odd number of errors. Since the errors on each qubit are independent, we can then take appropriate products of these probabilities to calculate $P(\ell_i|\ell_{i-1})$. The probability that the $k$th qubit will experience an even number of errors is
\begin{equation}\label{eq:poisson_even}
    P(e_k\text{ is even}) = \sum^{\infty}_{j = 0}P(e_k = 2j) = e^{-\mu T}\sum^{\infty}_{j = 0}\frac{(\mu T)^{2j}}{(2j)!} = e^{-\mu T}\cosh(\mu T),
\end{equation}
and therefore the probability of an odd number of errors is
\begin{equation}\label{eq:poisson_odd}
    P(e_k\text{ is odd}) = 1 - P(e_k\text{ is even}) = e^{-\mu T}\sinh(\mu T).
\end{equation}
Using Eq.~\eqref{eq:poisson_even} and Eq.~\eqref{eq:poisson_odd}, the transition probability $P(\ell_i|\ell_{i-1})$ is given by
\begin{equation}\label{eq:transition_prob}
    P(\ell_i|\ell_{i-1}) = [\sinh(\mu T)]^{d(\ell_i, \ell_{i-1})}[\cosh(\mu T)]^{3 - d(\ell_i, \ell_{i-1})}\exp[-3\mu T],
\end{equation}
where $d(\ell_i, \ell_{i-1})$ is the Hamming distance between the bit representation of $\ell_i$ and the bit representation of $\ell_{i - 1}$. In words, the probability consists of an exponential multiplied by a sinh term for every qubit that experiences a net flip and a cosh term for every qubit whose state is ultimately left unchanged.

Since the error rate $\mu$ is independent of time, the value of index $i$ is irrelevent to the value of $P(\ell_i|\ell_{i-1})$. This allows us to define a single $8 \times 8$ time-invariant transition matrix $J$ with elements $J_{\ell_{i-1}, \ell_i} \equiv P(\ell_i|\ell_{i-1})$. This matrix can be understood as the parameterization of a discrete Markov chain \cite{Norris_1997} which describes how bit-flip errors can alter the state of our system over time intervals of length $T$. For the remainder of this paper we will use the elements of $J$ to denote the transition probabilities $P(\ell_i|\ell_{i-1})$.

\subsection{Analysis of the measurement density}\label{sec:measure_distribution}

From our discussion of continuous measurement in Sec.~\ref{sec:measurement}, we know that the value of the $i$th measurement will depend on the average value of the syndrome from time $iT$ to $(i + 1)T$, plus an additive Gaussian noise term. Since it will be necessary to discuss measurements of the two parity operators separately, we unpack the vector quantity $\vec{M}_i$ as per Eq.~\eqref{eq:vec_measurement} into its components $M^{(1)}_i$ and $M^{(2)}_i$, which describe the results of measuring $Z_1Z_2$ and $Z_2Z_3$ respectively. The measurement distribution $P(\vec{M}_i|\ell_i\ell_{i-1})$ will be hereafter written as $P(M^{(1)}_iM^{(2)}_i|\ell_i\ell_{i-1})$ to make this distinction explicit. For convenience we denote the averaged syndrome values as $\bar{S}^{(1)}_i$ and $\bar{S}^{(2)}_i$, such that
\begin{equation}\label{eq:synd_bar_def}
    \bar{S}^{(j)}_i \equiv \frac{1}{T}\int^{T}_{0} S^{(j)}(iT + \tau)d\tau
\end{equation}
where $S^{(j)}(t)$ is the value of the $j$th syndrome at time $t$.

\begin{figure}
    \centering
    \includegraphics[width=\textwidth]{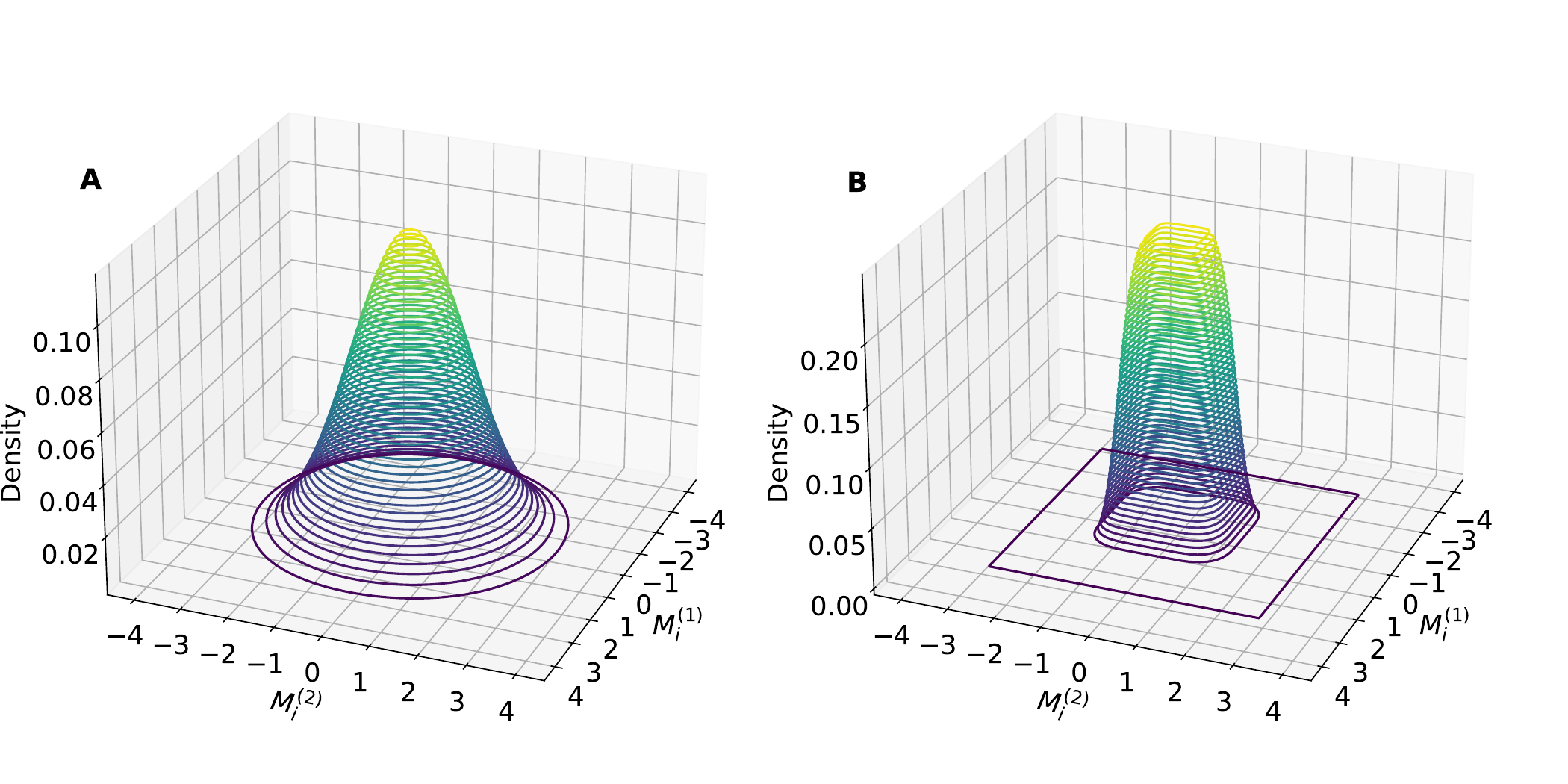}
    \caption{Contour plots of $P(M^{(1)}_iM^{(2)}_i|\ell_i\ell_{i-1})$ derived from Eq.~\eqref{eq:measurement_dist} using a uniform distribution over the syndrome means $\bar{S}^{(1)}_i$ and $\bar{S}^{(2)}_i$. Plot A was generated using a moderate amount of noise ($\frac{k}{T} = 1)$, and possesses a Gaussian-like shape. Plot B contains far less noise ($\frac{k}{T} = 0.05)$, so the square shape of the underlying uniform distribution starts to show through.}
    \label{fig:measure_density}
\end{figure}

Using this labeling, we have from Eq.~\eqref{eq:cont_avg} that $M^{(j)}_i \sim \mathcal{N}(\bar{S}^{(j)}_i, \frac{k}{T})$, so $P(M^{(1)}_iM^{(2)}_i|\ell_i\ell_{i-1})$ will broadly resemble a bivariate Gaussian distribution. However, since the means $\bar{S}^{(1)}_i$ and $\bar{S}^{(2)}_i$ are themselves random variables due to errors occurring within the integration period, the measurement distribution will consist of a Gaussian function integrated over mean values in the continuous interval [-1, +1]. This is given by
\begin{equation}\label{eq:measurement_dist}
    P(M^{(1)}_iM^{(2)}_i|\ell_i\ell_{i-1}) = \int^1_{-1}\int^1_{-1}\prod^2_{j=1}\frac{\exp[\frac{-T}{2k}(M^{(j)}_i - \Bar{S}^{(j)}_i)^2]}{\sqrt{2\pi\frac{k}{T}}} P(\bar{S}^{(1)}_i\bar{S}^{(2)}_i|\ell_i\ell_{i-1})d\bar{S}^{(1)}_id\bar{S}^{(2)}_i,
\end{equation}
where we assume that the measurement noise of $Z_1Z_2$ and $Z_2Z_3$ is uncorrelated and of equal variance. Parsing Eq.~\eqref{eq:measurement_dist}, we see that the measurement distribution has the form of a continuous Gaussian mixture, with each component centered at a different point in the 2D interval $[-1, +1] \times [-1, +1]$. The mixture components are weighted by the probability $P(\bar{S}^{(1)}_i\bar{S}^{(2)}_i|\ell_i\ell_{i-1})$ of observing those mean values, assuming that $\ell_i$ and $\ell_{i-1}$ are known. Figure~\ref{fig:measure_density} illustrates the Gaussian-like behavior of $P(M^{(1)}_iM^{(2)}_i|\ell_i\ell_{i-1})$, which is especially clear under strong noise. We exploit this similarity in our Bayesian algorithm to efficiently approximate the measurement log-likelihood as described in Sec.~\ref{sec:single_error}.

The fact that Eq.~\eqref{eq:measurement_dist} is not simply a bivariate Gaussian distribution with a fixed mean reflects the uncertainty that is inherent in our problem. Since $P(\bar{S}^{(1)}_i\bar{S}^{(2)}_i|\ell_i\ell_{i-1})$ is conditioned only on the state of the system at $iT$ and $(i + 1)T$, we must predict the values of the syndrome averages using this information alone. Figure~\hyperref[fig:syndromes]{1C} demonstrates the difficulty of such an inference, as the value of $\bar{S}^{(j)}_i$ depends not only on the precise \textit{number} of errors in the integration interval but more importantly on the \textit{locations} of these errors within the interval, all of which are unknown to us. There are an infinite number of possible values that $\bar{S}^{(1)}_i$ and $\bar{S}^{(2)}_i$ can take for any given $(\ell_{i-1}, \ell_{i})$ pair, so we must marginalize over these degrees of freedom in order to derive the measurement distribution.

To better understand where the probability of $\bar{S}^{(1)}_i$ and $\bar{S}^{(2)}_i$ is concentrated, we can condition $P(\bar{S}^{(1)}_i\bar{S}^{(2)}_i|\ell_i\ell_{i-1})$ on the number of errors occurring in the $i$th interval, which gives
\begin{equation}
\begin{split}\label{eq:synd_factorize}
    P(\bar{S}^{(1)}_i\bar{S}^{(2)}_i|\ell_i\ell_{i-1}) &= \sum^{\infty}_{e_1,e_2,e_3 = 0}P(\bar{S}^{(1)}_i\bar{S}^{(2)}_i|\ell_i\ell_{i-1};e_1e_2e_3)P(e_1e_2e_3|\ell_i\ell_{i-1})
    \\
    &= \sum^{\infty}_{e_1,e_2,e_3 = 0}P(\bar{S}^{(1)}_i\bar{S}^{(2)}_i|\ell_{i-1};e_1e_2e_3)P(e_1|\ell_i\ell_{i-1})P(e_2|\ell_i\ell_{i-1})P(e_3|\ell_i\ell_{i-1}),
\end{split}
\end{equation}
where $e_k$ is the number of errors occurring on the $k$th qubit during the $i$th interval (this time index is suppressed for notational convenience).  In the final line we exploit the fact that errors on different qubits are independent, and that $\ell_i$ is a deterministic function of $e_1$, $e_2$, $e_3$, $\ell_{i-i}$ and therefore does not need to be explicitly conditioned on.

From our discussion in Sec.~\ref{sec:markov}, we know that $\ell_i$ and $\ell_{i-1}$ only specify whether an even or odd number of errors occurred on each qubit, so $P(e_k|\ell_i\ell_{i-1})$ can be rewritten as $P(e_k|\text{odd})$ or $P(e_k|\text{even})$ depending on whether $\ell_i$ and $\ell_{i-1}$ differ in the $k$th qubit. These distributions are identical to the Poisson distribution from Eq.~\eqref{eq:poisson} except that the probability of either even or odd $e_k$ is set to zero. After re-normalizing, we get
\begin{equation}
    P(e_k|\text{odd}) =
    \begin{cases}
    \frac{(\mu T)^{e_k}}{e_k!\sinh(\mu T)} & e_k \text{ odd}
    \\
    0 & e_k \text{ even}
    \end{cases}
    , \quad \quad \quad P(e_k|\text{even}) =
    \begin{cases}
    \frac{(\mu T)^{e_k}}{e_k!\cosh(\mu T)} & e_k \text{ even}
    \\
    0 & e_k \text{ odd}
    \end{cases}
    .
\end{equation}
As an example of how these distributions are used, if $\ell_i = 4$ and $\ell_{i-1} = 0$ then the error distribution factorizes as $P(e_1e_2e_3|4,0) = P(e_1|\text{odd})P(e_2|\text{even})P(e_3|\text{even})$.

Referring back to Eq.~\eqref{eq:synd_factorize}, the final term left to characterize is $P(\bar{S}^{(1)}_i\bar{S}^{(2)}_i|\ell_{i-1};e_1e_2e_3)$, which describes how the syndrome averages are distributed given a starting state and the number of errors that occurred. This term is challenging to evaluate, since errors on the second qubit will flip both syndromes simultaneously and thus prevent the joint distribution from factorizing. If we consider errors on the second qubit separately from errors on the first and third qubits, i.e., focus on cases where $e_1, e_3 = 0$ and where $e_2 = 0$, then we can derive (see Appendix~\ref{sec_app:syndrome_dist}) the following expressions 
\begin{align}
    P(\bar{S}^{(1)}_i\bar{S}^{(2)}_i|\ell_{i-1};e_1 0 e_3) &= P(\bar{S}^{(1)}_i|\ell_{i-1};e_100)P(\bar{S}^{(2)}_i|\ell_{i-1};00e_3)
    \label{eq:no_second_error_factorize}
    \\[0.3cm]
    P(\bar{S}^{(1)}_i|\ell_{i-1};e_100) &= 
    \begin{cases}
        \delta(1 \mp_1 \Bar{S}^{(1)}_i) & e_1 = 0
        \\[0.2cm]
        \dfrac{e_1!}{2a_1!b_1!}\left(\dfrac{1 \pm_1 \Bar{S}^{(1)}_i}{2}\right)^{a_1}\left(\dfrac{1 \mp_1 \Bar{S}^{(1)}_i}{2}\right)^{b_1} & e_1 > 0
    \end{cases}
    \label{eq:no_second_error_marg_1}
    \\[0.3cm]
    P(\bar{S}^{(2)}_i|\ell_{i-1};00e_3) &= 
    \begin{cases}
        \delta(1 \mp_2 \Bar{S}^{(2)}_i) & e_3 = 0
        \\[0.2cm]
        \dfrac{e_3!}{2a_3!b_3!}\left(\dfrac{1 \pm_2 \Bar{S}^{(2)}_i}{2}\right)^{a_3}\left(\dfrac{1 \mp_2 \Bar{S}^{(2)}_i}{2}\right)^{b_3} & e_3 > 0
    \end{cases}
    \label{eq:no_second_error_marg_2}
    \\[0.3cm]
    P(\bar{S}^{(1)}_i\bar{S}^{(2)}_i|\ell_{i-1};0e_20) &= 
    \begin{cases}
        \delta(1 \mp_1 \Bar{S}^{(1)}_i)\delta(1 \mp_2 \Bar{S}^{(2)}_i) & e_2 = 0
        \\[0.2cm]
        \dfrac{e_2!}{2a_2!b_2!}\left(\dfrac{1 \pm_1 \Bar{S}^{(1)}_i}{2}\right)^{a_2}\left(\dfrac{1 \mp_1 \Bar{S}^{(1)}_i}{2}\right)^{b_2}\delta(\pm_1 \Bar{S}^{(1)}_i \mp_2 \bar{S}^{(2)}_i) & e_2 > 0
    \end{cases}
    \label{eq:no_second_error_marg_3}
\end{align}
where  $a_k$ and $b_k$ are the integer parts of $\frac{e_k}{2}$ and $\frac{e_k - 1}{2}$ respectively, and $\delta(x)$ is the Dirac delta function. The ``$\pm_j$'' term indicates whether state $\ell_{i-1}$ has even ``$+$'' or odd ``$-$'' parity with respect to the $j$th measurement operator, and therefore whether the syndrome should be added or subtracted. We note that the syndrome density is uniform (on its support) when only a single error occurs (i.e., $a_k = b_k = 0$). This simplification will be utilized in Sec.~\ref{sec:single_error} as we construct our logarithmic filter.   

\begin{figure}
    \centering
    \includegraphics[width=\textwidth]{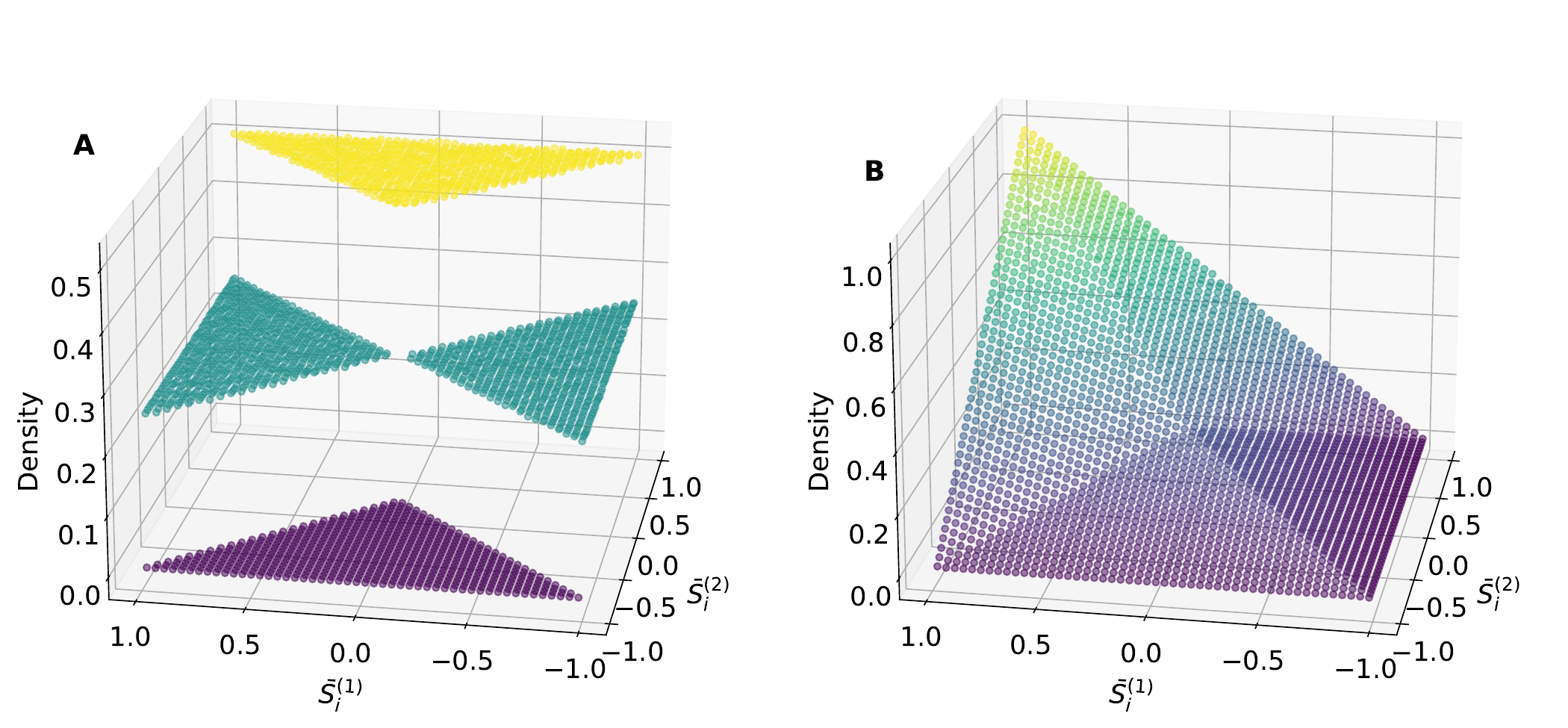}
    \caption{Plot A shows the probability density $P(\Bar{S}^{(1)}_i\bar{S}^{(2)}_i|0;011)$, which describes the syndrome means when the system starts in $\ket{000}$ and then experiences an error on the second and third qubits. Plot B shows $P(\Bar{S}^{(1)}_i\bar{S}^{(2)}_i|0;111)$, whose form is given in Eq.~\eqref{eq:111_example}. As a general rule, the more errors that occur in the interval the more ``smooth'' the syndrome density will appear, with plot A being outright discontinuous while plot B has a discontinuous first derivative. These plots were generated from $10^8$ samples using histograms with 50 equally spaced bins along each axis. The color gradient indicates the relative magnitude of each bin, with different scales for the two plots.}
    \label{fig:synd_densities}
\end{figure}

Unfortunately, when errors occur on the second qubit and on at least one of the other qubits, the joint distributions are non-smooth and appear to lack convenient analytic forms (see Appendix~\ref{sec_app:multi-qubit}). As an example, if the system starts in $\ket{000}$ and experiences one error on each qubit, then the syndrome distribution is given by
\begin{equation}\label{eq:111_example}
    P(\Bar{S}^{(1)}_i\bar{S}^{(2)}_i|0;111) = \frac{1}{4}\left[ \min(\Bar{S}^{(1)}_i, \Bar{S}^{(2)}_i) + \max(0, \Bar{S}^{(1)}_i + \Bar{S}^{(2)}_i)\right],
\end{equation}
which is a piecewise function with a discontinuous first derivative. Figure~\ref{fig:synd_densities} shows a plot of $ P(\Bar{S}^{(1)}_i\bar{S}^{(2)}_i|0;111)$, together with $P(\Bar{S}^{(1)}_i\bar{S}^{(2)}_i|0;011)$ which is even less smooth. 

\sloppy Due to the poor behavior of the syndrome densities, we are unable to make further progress toward an exact, analytic expression for the measurement probability $P(M^{(1)}_iM^{(2)}_i|\ell_i\ell_{i-1})$. Most of this difficulty stems from the fact that we must, in general, account for an infinite number of possible errors that could occur across all three qubits within a given integration interval. If, however, we restrict ourselves to situations where only a single error occurs, which is reasonable when $\mu$ and $T$ are small, then the expressions in Eqs.~(\ref{eq:no_second_error_marg_1}~--~\ref{eq:no_second_error_marg_3}) can be used directly and we will be able to evaluate the integrals in Eq.~\eqref{eq:measurement_dist}. This is the approach we take in Sec.~\ref{sec:single_error}, although it is no longer an exact treatment of the problem.  

\subsection{Constructing the Bayesian Filter}

Although Sec.~\ref{sec:measure_distribution} suggests that the posterior distribution for the state tracking problem does not possess a convenient functional form, it is still useful to synthesize our results from this section into an algebraic description of how the optimal model would operate. Referring back to Eq.~\eqref{eq:bayes_deriv}, the posterior state probabilities $\hat{P}(\ell_i)$ are given recursively in terms of the measurement probabilities $P({M}^{(1)}_iM^{(2)}_i|\ell_i\ell_{i-1})$ and transition probabilities $J_{\ell_{i-1}\ell_i} \equiv P(\ell_i|\ell_{i-1})$. Noting the dependence of both terms on $\ell_i$ and $\ell_{i-1}$, the update rule is given by
\begin{equation}\label{eq:bayes_filter}
\begin{split}
     \hat{P}(\ell_i) &\propto \sum^7_{\ell_{i-1} = 0} \hat{P}(\ell_{i-1})P(M^{(1)}_iM^{(2)}_i|\ell_i\ell_{i-1})J_{\ell_{i-1}\ell_i}
     \\
     &\propto \sum^7_{\ell_{i-1} = 0}\hat{P}(\ell_{i-1})\Tilde{J}(M^{(1)}_iM^{(2)}_i)_{\ell_{i-1}\ell_i},
\end{split}
\end{equation}
where $\Tilde{J}(M^{(1)}_iM^{(2)}_i)_{\ell_{i-1}\ell_i} \equiv P(M^{(1)}_iM^{(2)}_i|\ell_i\ell_{i-1})J_{\ell_{i-1}\ell_i}$ can be viewed as a weighted transition matrix which combines the Markov dynamics of the error process with the likelihood of the observed measurement outcomes.

Due to the recursive nature of Eq.~\eqref{eq:bayes_filter}, the state tracking model operates naturally as a digital filter, taking in syndrome measurements at each time step and outputting a set of posterior probabilities that incorporate all of the information available to us. Setting aside the challenges of computing $P({M}^{(1)}_iM^{(2)}_i|\ell_i\ell_{i-1})$ from Eq.~(\ref{eq:measurement_dist}), this filter is optimal in the sense that it was constructed from the formal probability manipulations in Eq.~\eqref{eq:bayes_deriv}, which assumed only that the measurements and errors were Markovian. In Sec.~\ref{sec:measure_distribution} we made further assumptions about the structure of the measurement signal, and insofar as these assumptions are valid there is nowhere for the algorithm to be improved. Of course, should the physical system not obey the idealized model outlined in Sec.~\ref{sec:measurement}, then the filter will need to be modified in order to remain optimal.

From a computational standpoint, Eq.~\eqref{eq:bayes_filter} represents the filter as a simple matrix-vector product, with the difficulty centered on calculating $\tilde{J}$. More specifically, Sec.~\ref{sec:measure_distribution} made it clear that computing $P({M}^{(1)}_iM^{(2)}_i|\ell_i\ell_{i-1})$ requires us to evaluate the integral in Eq.~\eqref{eq:measurement_dist}, which appears to be analytically intractable when the error combinations are not restricted. This limitation, together with other issues that will be discussed in Sec.~\ref{sec:obstacles}, rules out using the optimal filter for real-time error tracking. That said, accurate numerical evaluation of the integral is still possible via sampling techniques, which are discussed further in Appendix~\ref{sec_app:implement_filter}, so we will use this filter as a benchmark for other state tracking methods.

\section{Obstacles and Prior Work}\label{sec:obstacles}

In the context of real-time quantum error tracking, where latency is measured in nanoseconds and hardware resources can be tight, the Bayesian filter outlined in Sec.~\ref{sec:optimal} has four significant challenges:
\begin{enumerate}
    \item The Gaussian integral in  Eq.~\eqref{eq:measurement_dist} cannot be evaluated analytically, and sampling methods are too resource-intensive for real-time filtering.
    
    \item Computing even a single Gaussian requires the implementation of an exponential function, which can be challenging on low-latency, high-throughput devices such as FPGAs \cite{Pang_Membrey_2017}.
    
    \item The outputs of the Gaussians will likely need to be represented using floating-point numbers to capture the necessary range and precision, which adds a further computational burden.
    
    \item The probabilities generated by Eq.~\eqref{eq:bayes_filter} will need to be periodically normalized in order to prevent overflow or underflow, which requires a dedicated division routine that will take up resources and be slow to run.
\end{enumerate}
Overcoming all of these obstacles while preserving the underlying Bayesian framework is not trivial.

The first derivation of an optimal continuous-time filter for the three qubit bit-flip code was published by van Handel and Mabuchi, who recovered the well-known Wonham filter after solving for the least-squares estimator of the density matrix \cite{Mabuchi_2009}. This classical filter was designed to output probabilities for the states of a Markov chain observed continuously by a signal with additive Gaussian noise. Using a modified version of our notation, the canonical Wonham filter has the form
\begin{equation}\label{eq:wonham_filter}
    d \Hat{P}_{\ell}(t) = \sum^7_{\ell' = 0} \Hat{P}_{\ell'}(t)Q_{\ell'\ell}dt + \frac{1}{k}\sum^2_{j=1}[S^{(j)}_{\ell} - \bar{\bar{S}}^{(j)}(t)]\Hat{P}_{\ell}(t)[dM^{(j)}(t) -  \bar{\bar{S}}^{(j)}(t)dt],
\end{equation}
where $\Hat{P}_\ell(t)$ is the probability of the $\ell$th state at time $t$, $S^{(j)}_{\ell} \in \{-1, +1\}$ is the syndrome of the $\ell$th state from the $j$th parity operator, $\bar{\bar{S}}^{(j)}(t) \equiv \sum_{\ell}\hat{P}_\ell(t) S^{(j)}_\ell$ is the average of the $j$th syndrome at time $t$, and $Q = \frac{1}{T}\ln J$ is the rate matrix of the continuous Markov chain which describes the state transitions.  

It is important to emphasize that Eq.~\eqref{eq:bayes_filter} and Eq.~\eqref{eq:wonham_filter} are each optimal with respect to different measurement regimes. The Wonham filter is optimal under the assumption that we have access to instantaneous signal readouts at arbitrary $t$, while our filter in Eq.~\eqref{eq:bayes_filter} is optimal when the measurements are restricted to being integrated averages of the signal over some finite period $T$ as in Eq.~\eqref{eq:cont_avg}. As $T \rightarrow 0$ these two forms of measurement converge, and thus Eq.~\eqref{eq:bayes_filter} converges to Eq.~\eqref{eq:wonham_filter} after normalization (see Appendix~\ref{sec_app:converge}). The general problem of filtering Markov jump processes using discrete observations has recently been explored by Borisov \cite{Borisov_2020}, who derives results that are in agreement with our work in Sec.~\ref{sec:optimal}.

As a non-linear stochastic differential equation, the Wonham filter is not practical to use in its exact form, though it is still possible to create discretized approximations of Eq.~\eqref{eq:wonham_filter} which are effective. For example, Gange George et al.\ applied the Euler-Maruyama method to a logarithmic transformation of the Wonham filter in order to derive a first-order update rule that was numerically stable \cite{Gang_George_Zhang_Liu_2004}. In the specific context of continous error correction, Mohseninia et al.\ \cite{Mohseninia_Yang_Siddiqi_Jordan_Dressel_2020} proposed a linearized version of Eq.~\eqref{eq:wonham_filter} equivalent to
\begin{equation}\label{eq:linear_wonham}
    \hat{P}(\ell_i) \propto \hat{P}(\ell_{i-1} = \ell_i) + T\sum^7_{\ell_{i-1} = 0}\left[Q_{\ell_{i-1}\ell_i} + \frac{M^{(1)}_iS^{(1)}_{\ell_{i-1}} + M^{(2)}_iS^{(2)}_{\ell_{i-1}}}{k}\delta_{\ell_{i-1}\ell_i}\right]\hat{P}({\ell_{i-1}}),
\end{equation}
where $S^{(j)}_{\ell_{i-1}}$ and $Q$ are defined as in Eq.~\eqref{eq:wonham_filter}. The filter described by Eq.~\eqref{eq:linear_wonham} does not involve any Gaussian functions and therefore avoids the first three obstacles in our list, but still requires periodic normalization. Additionally, as $T$ grows larger the accuracy of this first-order approximation will decline.

The filters discussed so far have all been motivated at some level by Bayesian probability analysis, but there exists another class of algorithm, referred to as \textit{threshold} or \textit{boxcar} filters, which eschews connections to probability theory in favor of computational simplicity \cite{Hieftje_1972}. At a basic level, these models take a pair of integrated measurement values $(M^{(1)}_i, M^{(2)}_i)$ and compare them to a set of predefined thresholds, after which an appropriate action is taken. Atalaya et al.\ developed a double-threshold algorithm for the Bacond-Shor code using non-commuting observables \cite{Atalaya_Bahrami_Pryadko_Korotkov_2017}, while Mosheninia et al.\ proposed a non-Markovian boxcar filter and a double threshold scheme for the three-qubit repetition code which were easy to implement and attained good performance \cite{Mohseninia_Yang_Siddiqi_Jordan_Dressel_2020}. Atalaya and Zhang et al.\ applied a flexible double thresholding scheme to a system undergoing quantum annealing, demonstrating that these algorithms were effective at error correction even in the presence of Hamiltonian evolution \cite{Atalaya_Zhang_Niu_Babakhani_Chan_Epstein_Whaley_2021}.

\section{A Practical Bayesian Filter}\label{sec:practical_filter}

To address the obstacles identified in Sec.~\ref{sec:obstacles}, we propose an algorithm that avoids any exponentiation or division operations, while also approximating the analytically intractable integral in Eq.~\eqref{eq:measurement_dist}. In Sec.~\ref{sec:single_error} we simplify Eq.~\eqref{eq:measurement_dist} by assuming that only a single error occurs during each integration period, which yields simple expressions for $P(\bar{S}^{(1)}_i\bar{S}^{(2)}_i|\ell_i\ell_{i-1})$. In Sec.~\ref{sec:log_prob} we eliminate the need for exponentiation and division by shifting our analysis into log-probability space, where the multiplication of probabilities becomes addition, normalization becomes subtraction, and Gaussian distributions transform into simple quadratic functions. Finally, Sec.~\ref{sec:term_strategies} introduces what we call ``single-term'' and ``two-term'' strategies for approximating the LogSumExp functions needed for Markov chain evolution using log-probabilities. The performance of our algorithm is tested numerically in Sec.~\ref{sec:numerical}.

\subsection{Single-error approximation}\label{sec:single_error}

The largest obstacle to using Eq.~\eqref{eq:bayes_filter} as a filter is that $P(M^{(1)}M^{(2)}|\ell_i\ell_{i-1})$ cannot be easily evaluated. As explored in Sec.~\ref{sec:measure_distribution}, this difficulty arises because the syndrome density terms $P(\bar{S}^{(1)}_i\bar{S}^{(2)}_i|\ell_{i-1};e_1e_2e_3)$ do not in general have convenient analytic forms, especially when the number of errors is large. This in turn prevents us from solving the integral in Eq.~\eqref{eq:measurement_dist}. However, if we were to assume that only a single error can occur within a given measurement interval, then Eqs.~(\ref{eq:no_second_error_marg_1} -- \ref{eq:no_second_error_marg_3}) would reduce to
\begin{align}
    P(\bar{S}^{(1)}_i|\ell_{i-1}; e_1 0 0) &=
    \begin{cases}
        \delta(1 \mp_1 \bar{S}^{(1)}) & e_1 = 0
        \\[0.1cm]
        \frac{1}{2} & e_1 = 1
    \end{cases}
    \label{eq:single_error_1}
    \\[0.2cm]
    P(\bar{S}^{(1)}_i|\ell_{i-1}; 0 0 e_3) &=
    \begin{cases}
        \delta(1 \mp_2 \bar{S}^{(2)}_i) & e_3 = 0
        \\[0.1cm]
        \frac{1}{2} & e_3 = 1
    \end{cases}
    \label{eq:single_error_3}
    \\[0.2cm]
    P(\bar{S}^{(1)}_i\bar{S}^{(2)}_i|\ell_{i-1}; 0 e_2 0) &=
    \begin{cases}
        \delta(1\mp_1 \bar{S}^{(1)}_i)\delta(1 \mp_2 \bar{S}^{(2)}_i) & e_2 = 0
        \\[0.1cm]
        \frac{1}{2}\delta(\pm_1\bar{S}^{(1)}_i \mp_2 \bar{S}^{(2)}_i) & e_2 = 1,
    \end{cases}
    \label{eq:single_error_2}
\end{align}
which all lead to tractable integrals when substituted into Eq.~\eqref{eq:measurement_dist}. This single-error approximation, which is reasonable when the average number of errors per interval, $\mu T$, is small, leads to Gaussian-like measurement distributions which can be easily incorporated into our log-probability filter.

For simplicity, we first consider an error that occurs on either the first ($e_1 = 1)$ or third ($e_3 = 1)$ qubit, since it will affect only one of the syndromes and thus allow the joint distribution to separate as in Eq.~\eqref{eq:no_second_error_factorize}. By substituting $P(\bar{S}^{(1)}_i\bar{S}^{(2)}_i|\ell_{i-1}; 100)$ into Eq.~\eqref{eq:measurement_dist} and requiring that $\ell_i$ and $\ell_{i-1}$ differ only in the first qubit, we get
\begin{equation}\label{eq:single_error_deriv}
    P(M^{(1)}_iM^{(2)}_i|\ell_{i-1};100) = \frac{1}{4}\left[\text{erf}\left(\frac{M^{(1)}_i + 1}{\frac{2k}{T}}\right) - \text{erf}\left(\frac{M^{(1)}_i - 1}{\frac{2k}{T}}\right)\right]\cdot \frac{\exp[\frac{-T}{2k}(M^{(2)}_i \mp_2 1)^2]}{\sqrt{2\pi\frac{k}{T}}},
\end{equation}
where $\text{erf}(x) \equiv \frac{2}{\sqrt{\pi}}\int^x_0 e^{-y^2}dy$ is the error function. The last line of Eq.~\eqref{eq:single_error_deriv} consists of two distinct terms, one of which is a function of $M^{(1)}_i$ and the other of $M^{(2)}_i$. The $M^{(2)}_i$ term is simply a Gaussian centered at $\pm_2 1$, since the second syndrome is not sensitive to an error on the first qubit. The $M^{(1)}_i$ term, by contrast, arises because the mean of the first syndrome is distributed uniformly over the range $[-1, +1]$, which yields a pair of error functions. Notably, this term does not depend on $\pm_1$, so it is independent of the state of the first two qubits.

While the first term in Eq.~\eqref{eq:single_error_deriv} is an exact result from of our single-error integration, it is not very convenient to evaluate numerically. As described in Appendix~\ref{sec_app:gauss_fit}, the measurement distribution can be approximated as
\begin{equation}\label{eq:erf_approx_density}
    P(M^{(1)}_iM^{(2)}_i|\ell_{i-1};100) = \frac{\exp[\frac{-3}{2(1 + 3\frac{k}{T})}(M^{(1)}_i)^2]}{\sqrt{2\pi(\frac{1}{3} + \frac{k}{T}})} \cdot \frac{\exp[\frac{-T}{2k}(M^{(2)}_i \mp_2 1)^2]}{\sqrt{2\pi\frac{k}{T}}},
\end{equation}
which will be easy to evaluate numerically after we shift to log-probabilities in Sec.~\ref{sec:log_prob}. The measurement density for an error on the third qubit can be obtained from Eq.~\eqref{eq:erf_approx_density} by simply swapping $M^{(1)}_i$ and $M^{(2)}_i$ and replacing $\mp_2$ with $\mp_1$. Carrying out a similar procedure on the second qubit gives 
\begin{equation}\label{eq:second_qubit_approx}
    P(M^{(1)}_iM^{(2)}_i|\ell_{i-1};010) = \frac{1}{2}\frac{\exp[\frac{-T}{k}(\frac{M^{(1)}_i \mp_c M^{(2)}_i}{2})^2]}{\sqrt{\pi\frac{k}{T}}}\cdot\frac{\exp[\frac{-3}{2(1 + 3\frac{k}{2T})}(\frac{M^{(1)}_i \pm_c M^{(2)}_i}{2})^2]}{\sqrt{2\pi(\frac{1}{3} + \frac{k}{2T}})},
\end{equation}
where $\pm_c$ is positive when $\ell_{i-1}$ has the same parity with respect to $Z_1Z_2$ and $Z_2Z_3$ but negative when the parities are opposite (roughly, ``$\pm_c \equiv \pm_1\cdot\pm_2$'').

The final case to consider is when no errors occur in an interval at all. With reasonable values of $\mu$ and $T$ this is the most likely outcome for any given interval, and it's measurement distribution can be solved for easily by substituting $P(\bar{S}^{(1)}_i\bar{S}^{(2)}_i|\ell_i\ell_{i-1}) = P(\bar{S}^{(1)}_i|\ell_{i-1};000)P(\bar{S}^{(2)}_i|\ell_{i-1};000)$ into Eq.~\eqref{eq:measurement_dist}. This gives
\begin{equation}\label{eq:no_error_deriv}
    P(M^{(1)}_iM^{(2)}_i|\ell_{i-1};000) = \frac{\exp[\frac{-T}{2k}(M^{(1)}_i \mp_1 1)^2]}{\sqrt{2\pi\frac{k}{T}}}\cdot\frac{\exp[\frac{-T}{2k}(M^{(2)}_i \mp_2 1)^2]}{\sqrt{2\pi\frac{k}{T}}},
\end{equation}
which is simply the product of Gaussian distributions with means corresponding to the parities of $\ell_{i-1}$. 

Taken together, Eqs.~(\ref{eq:erf_approx_density}, \ref{eq:second_qubit_approx}, \ref{eq:no_error_deriv}) constitute the approximate description of $P(M^{(1)}_iM^{(2)}_i|\ell_i\ell_{i-1})$ that we will use to construct our log-probability filter. The Gaussian form of each equation is especially important on a practical level, as the log-densities will all reduce to sums of simple quadratic equations that can be easily computed.

\subsection{Moving to logarithmic probability}\label{sec:log_prob}

To begin the transformation from probabilities to log-probabilities, we take the logarithm of the optimal Bayesian filter from Eq.~\eqref{eq:bayes_filter}
\begin{equation}\label{eq:first_log}
\begin{split}
    \log \hat{P}(\ell_i) &= \log \sum^7_{\ell_{i-1} = 0} \hat{P}(\ell_{i-1})P(M^{(1)}_iM^{(2)}_i|\ell_i\ell_{i-1})J_{\ell_{i-1}\ell_i}
     \\
     &= \log \sum^7_{\ell_{i-1} = 0}\exp[\log \hat{P}(\ell_{i-1}) + \log \Tilde{J}(M^{(1)}_iM^{(2)}_i)_{\ell_{i-1}\ell_i}],
\end{split}
\end{equation}
where in the last line we exponentiate the logarithm of each term in the sum in order to preserve the recursive structure of the filter. Since $
\log \tilde{J}_{\ell_{i-1}\ell_i} \equiv \log J_{\ell_{i-1}\ell_i} + \log P(M^{(1)}_iM^{(2)}_i|\ell_i\ell_{i-1})$, the terms in Eq.~\eqref{eq:first_log} can be evaluated using our results from Sec.~\ref{sec:markov} and Sec.~\ref{sec:single_error}. 

The transition log-probability is given by the logarithm of Eq.~\eqref{eq:transition_prob}, which yields
\begin{equation}\label{eq:log_transitions}
    \log J_{\ell_{i-1}\ell_i} = 
     d(\ell_i, \ell_{i-1})\log \sinh(\mu T) + [3 - d(\ell_i, \ell_{i-1})]\log \cosh(\mu T) - 3\mu T
\end{equation}
where $d(\ell_i, \ell_{i-1})$ is the Hamming distance between $\ell_i$ and $\ell_{i-1}$. While the form of Eq.~\eqref{eq:log_transitions} is not especially illuminating, the value of $\mu T$ will be known to us and assumed to be constant across an experiment. We are therefore able to compute the values of $\log J_{\ell_{i-1}\ell_i}$ in advance and use them in the filter without needing to evaluate the $\log \sinh$ or $\log \cosh$ functions in real time. 

The measurement log-probabilities, by contrast, must be calculated anew each time a measurement is recorded. Such rapid computation is feasible because Eqs.~(\ref{eq:erf_approx_density}, \ref{eq:second_qubit_approx}, \ref{eq:no_error_deriv}) are all in Gaussian form, and therefore have the following quadratic log-probabilities
\begin{align}
    \log P(M^{(1)}_iM^{(2)}_i|\ell_{i-1}; 000) &= -\frac{T}{2k}[(M^{(1)}_i \mp_1 1)^2 - (M^{(2)}_i \mp_2 1)^2] + \log N
    \label{eq:log_000}
    \\[0.1cm]
    \log P(M^{(1)}_iM^{(2)}_i|\ell_{i-1}; 100) &= -\frac{3}{2(1+3\frac{k}{T})}(M^{(1)}_i)^2 - \frac{T}{2k}(M^{(2)}_i \mp_2 1)^2 + \log N'
    \\[0.1cm]
    \log P(M^{(1)}_iM^{(2)}_i|\ell_{i-1}; 010) &= -\frac{T}{k}(\frac{M^{(1)}_i \mp_c M^{(2)}_i}{2})^2 - \frac{3}{2(1 + 3\frac{k}{2T})}(\frac{M^{(1)}_i \pm_c M^{(2)}_i}{2})^2 + \log N''
    \\[0.1cm]
    \log P(M^{(1)}_iM^{(2)}_i|\ell_{i-1}; 001) &= -\frac{T}{2k}(M^{(1)}_i \mp_1 1)^2 - \frac{3}{2(1+3\frac{k}{T})}(M^{(2)}_i)^2 + \log N''',
    \label{eq:log_001}
\end{align}
where the normalization constants have been left unspecified for convenience. The different variances, all functions of $k$ and $T$, are constant during an experiment and therefore do not need to be computed at each step.

\begin{figure}
    \centering
    \includegraphics[width=0.7\textwidth]{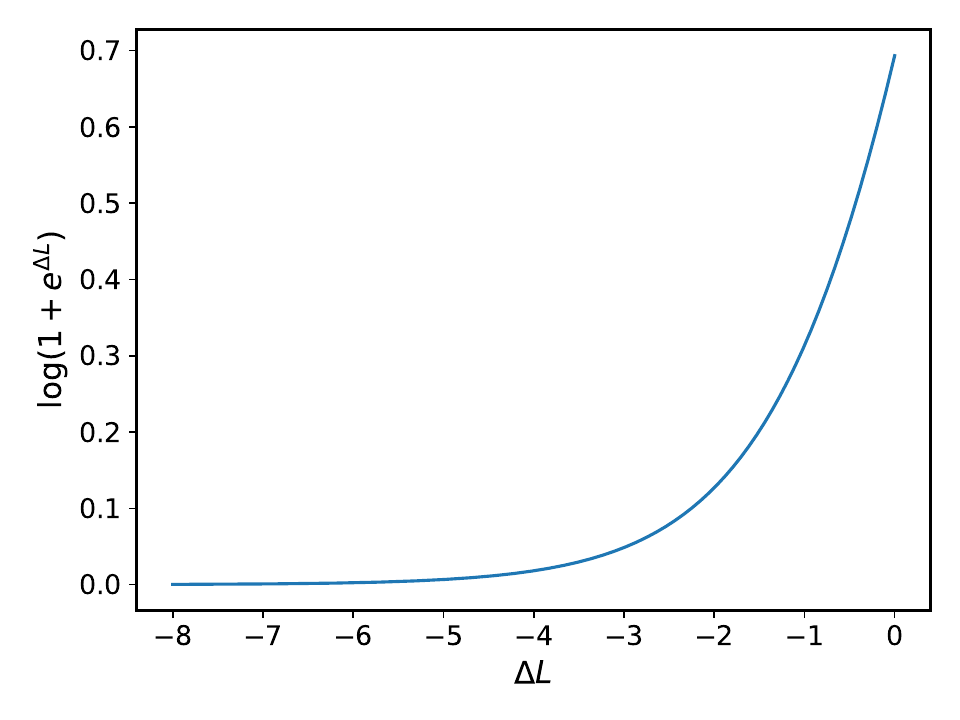}
    \caption{Plot of $\log(1 + e^{\Delta L})$, which goes to zero as the magnitude of $\Delta L \equiv L^{\ell_i}_{\ell_{i-1}} - L^{\ell_i}_*$ increases. This implies that $L^{\ell_i}_* + \log(1 + e^{\Delta L})$ rapidly approaches $L^{\ell_i}_*$ as the gap between $L^{\ell_i}_*$ and the other terms increases. Note that $\Delta L$ is never positive, since $L^{\ell_i}_*$ is the largest of all $L^{\ell_i}_{\ell_{i-1}}$ by definition.}
    \label{fig:logsumexp}
\end{figure}

Using Eqs.~(\ref{eq:log_transitions} -- \ref{eq:log_001}), we can calculate $\log \tilde{J}$ and therefore evaluate the exponent of each term in the sum of Eq.~\eqref{eq:first_log}. The last line of Eq.~\eqref{eq:first_log} is known as a \textit{LogSumExp} \cite{Blanchard_Higham_Higham_2020}, and its evaluation is the final obstacle to implementing our filter. For convenience we adopt the more compact notation $L^{\ell_i}_{\ell_{i-1}} \equiv \log \tilde{J}_{\ell_{i-1}\ell_i} + 
\log \hat{P}(\ell_{i-1})$, which serve as the inputs of the LogSumExp function,
\begin{equation}\label{eq:logsumexp}
    \log \hat{P}(\ell_i) = \log \sum^7_{\ell_{i-1} = 0}\exp[ L^{\ell_i}_{\ell_{i-1}}].
\end{equation}
Using this form, we exploit a well-known trick for evaluating LogSumExp functions by pulling the largest exponential outside of the sum
\begin{equation}\label{eq:max_logsumexp}
   \log \sum^7_{\ell_{i-1} = 0}\exp[ L^{\ell_i}_{\ell_{i-1}}] = L^{\ell_i}_* + \log(1 +  \sum_{\ell_{i-1} \neq *}\exp[L^{\ell_i}_{\ell_{i-1}} - L^{\ell_i}_*]),
\end{equation}
where $L^{\ell_i}_*$ is the largest input to the LogSumExp function. The second term in Eq.~\eqref{eq:max_logsumexp} is now the logarithm of one plus a set of terms whose exponents are all non-positive. Figure~\ref{fig:logsumexp} shows that as the difference between $L^{\ell_i}_*$ and the other inputs grows, the output of Eq.~\eqref{eq:max_logsumexp} will converge to $L^{\ell_i}_*$. If the difference is large for some subset of the inputs, then the expression can be simplified by removing that subset from the sum.

\subsection{Evaluating the LogSumExp}\label{sec:logsumexp_approximation}

To take advantage of Eq.~\eqref{eq:max_logsumexp}, we must know the relative magnitudes of the various $L^{\ell_i}_{\ell_{i-1}}$ terms in a typical run. For convenience, we introduce the non-positive quantity $\Delta L^{\ell_i}_{\ell_{i-1}} \equiv L^{\ell_i}_{\ell_{i-1}} - L^{\ell_i}_*$, which is simplified to just $\Delta L$ when the specific indices are not relevant. Using Figure~\ref{fig:logsumexp} as a guide, it is clear that when 
\begin{equation}\label{eq:threshold}
    \Delta L \leq -4 
\end{equation}
the magnitude of $\log(1 + e^{\Delta L})$ is negligible. This provides a threshold to determine whether a given $L^{\ell_i}_{\ell_{i-1}}$ contributes significantly to Eq.~\eqref{eq:max_logsumexp} or if it can instead be reasonably ignored.

\begin{figure}
    \centering
    \includegraphics[width=\textwidth]{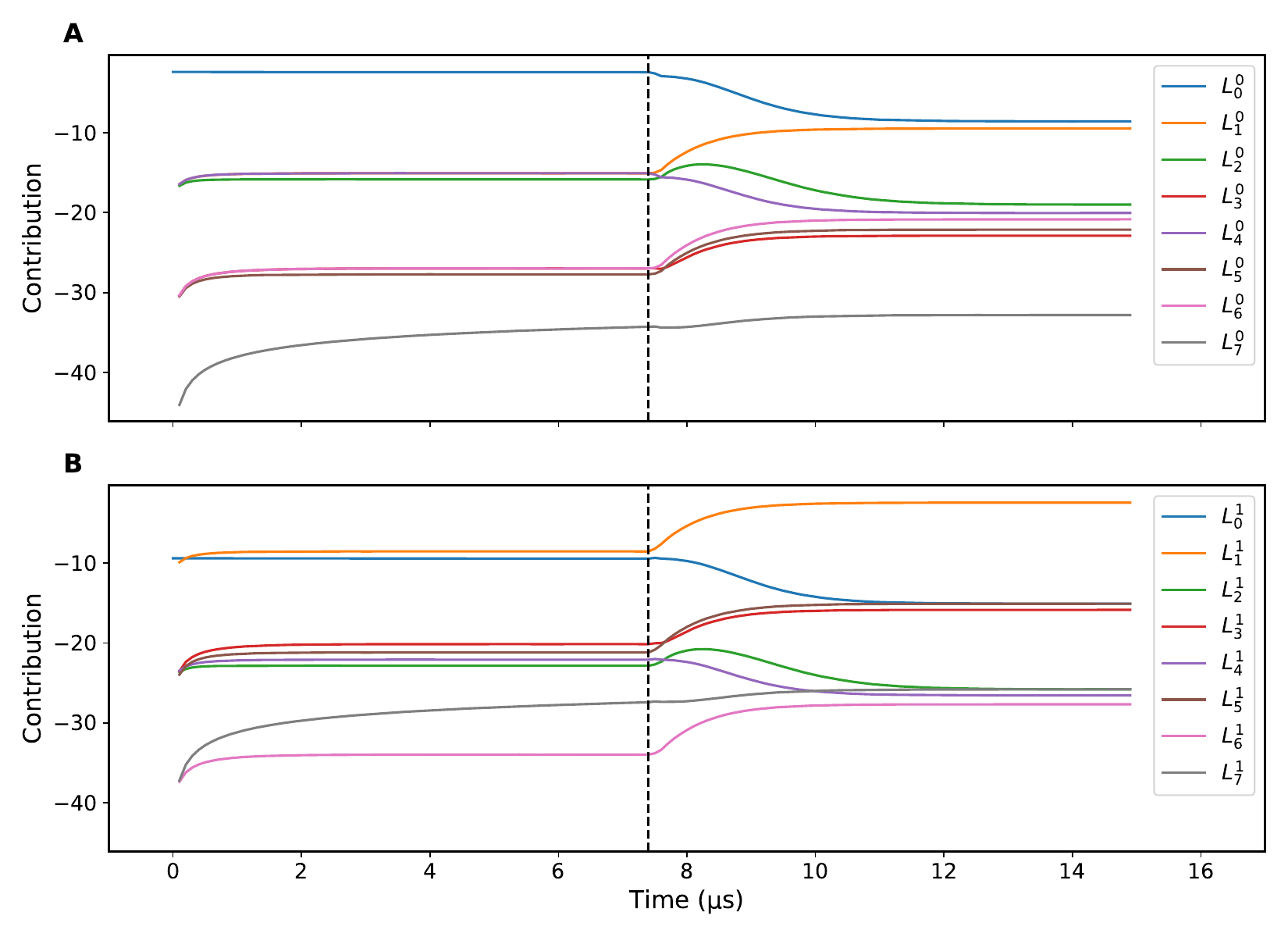}
    \caption{Plots of average $L^{\ell_i}_{\ell_{i-1}}$ contributions to the exponent of Eq.~\eqref{eq:max_logsumexp} for $\ket{000}$ (plot A) and $\ket{001}$ (plot B) taken from the optimal filter during a 15 $\mu s$ run, with $\frac{k}{T} = 4$ and a state that was initialized to $\ket{000}$. The dashed black lines mark the occurrence of a bit-flip error at $7.5\ \mu$s on the third qubit, which takes the system to $\ket{001}$. The plotted data represent averages over the measurement noise taken from $10^5$ different runs. When the system is $\ket{000}$, plot A shows clearly that only $L^0_0$ (blue line) is significant, while plot B shows that both $L^1_0$ (blue line) and $L^1_1$ (orange line) are significant. After the bit-flip occurs the state is $\ket{001}$ and the situation is reversed, with $L^1_1$, $L^0_0$, and $L^0_1$ contributing significantly.}
    \label{fig:contributions}
\end{figure}

In Figure~\ref{fig:contributions}, we plot the average values of $L^{\ell_i}_{\ell_{i-1}}$ for $\ell_i = 0$ and $\ell_i = 1$ before and after a bit-flip error occurs on a state initialized to $\ket{000}$. These values are taken from the optimal filter in Eq.~\eqref{eq:bayes_filter}, which we simulate numerically using the method described in Appendix~\ref{sec_app:implement_filter}. The plots of Figure~\ref{fig:contributions} demonstrate clearly that only one $L^{\ell_i}_{\ell_{i-1}}$ term is significant when $\ell_i$ is equal to the true system state and two of the terms are significant when $\ell_i$ is separated from the true state by one bit-flip. For example, the gap $\Delta L$ between $L^0_0$ and the second-largest $L^{\ell_i}_{\ell_{i-1}}$ in plot \hyperref[fig:contributions]{5A} is roughly -15 before the bit-flip, which is far smaller than the cutoff threshold of -4 described in Eq.~\eqref{eq:threshold}. Plots of the other $\ell_i$ indicate that states separated from the true state by two bit-flips have four significant terms and the state separated by three bit-flips has eight significant terms.

To understand the behavior shown in Figure~\ref{fig:contributions}, it is necessary to determine the typical magnitudes of Eq.~\eqref{eq:log_transitions} and Eqs.(\ref{eq:log_000} -- \ref{eq:log_001}) under realistic conditions. Starting first with the measurement log-densities, we are interested in their expected values when the system is in some state $\ell'$, so we take the average of Eqs.(\ref{eq:log_000} -- \ref{eq:log_001}) with respect to measurements distributed as $P(M^{(1)}_iM^{(2)}_i|\ell'\ell')$. For convenience we label this quantity as
\begin{equation}
    G^{\ell_i}_{\ell_{i-1}} \equiv E[\log P(M^{(1)}_iM^{(2)}_i|\ell_i\ell_{i-1})],
\end{equation}
and after some algebra the average log-densities are shown to scale as
\begin{equation}\label{eq:measurement_ratio}
    G^{\ell_i}_{\ell_{i-1}} = 
    \begin{cases}
        \mathcal{O}(-\log\frac{T}{k}) & \frac{T}{k} \ll 1,\ \text{all } \ell_i, \ell_{i-1}
        \\
        \mathcal{O}(-\log\frac{T}{k}) & \frac{T}{k} \gg 1,\ \ell_{i} \text{ or }\ell_{i-1} \in \{\ell', \ell'\oplus 7\}
        \\
        \mathcal{O}(-\frac{T}{k}) & \frac{T}{k} \gg 1,\ \ell_{i} \text{ or }\ell_{i-1} \notin \{\ell', \ell'\oplus 7\},
    \end{cases}
\end{equation}
where $\oplus$ is the bitwise exclusive-or and $\frac{T}{k}$ is the \textit{precision} of the noise (reciprocal of its variance). In words, Eq.~\eqref{eq:measurement_ratio} states that the log-densities all have a similar scale when the noise is large ($\frac{T}{k} \ll 1$) but begin to diverge when the noise is small ($\frac{T}{k} \gg 1$), with transitions to/from $\ell'$ and its complement having a significantly greater likelihood. This is expected, as the measurement densities differ primarily in the locations of their mean values, which become more obvious as the widths of the distributions shrink. With respect to what is shown in Figure~\ref{fig:contributions}, Eq.~\eqref{eq:measurement_ratio} indicates that unless the noise magnitude is very small, the $L^{\ell_i}_{\ell_{i-1}}$ will not differ greatly from one another based on their $G^{\ell_i}_{\ell_{i-1}}$ values.

Turning our attention to the transition probabilities, we know from the numerical tests in Sec.~\ref{sec:error_plots} that when $\mu T$ is on the order of $10^{-2}$ it becomes impossible to accurately track the state of the system at moderate run times, even for the optimal filter. We can therefore expand Eq.~\eqref{eq:log_transitions} for small $\mu T$ as
\begin{equation}\label{eq:transition_log_expansion}
    \log J_{\ell_{i-1}\ell_i} =
        d(\ell_i, \ell_{i-1})\log(\mu T) + \mathcal{O}(\mu T),
\end{equation}
and keep only the leading $\mu T$ term. For fixed $\mu T$, the values of $\log J_{\ell_{i-1}\ell_i}$ differ only in how many factors of $\log(\mu T)$ are included based on the Hamming distance $d(\ell_i, \ell_{i-1})$. Since $\log(10^{-2}) \approx -4.6$, the magnitude of a given $L^{\ell_i}_{\ell_{i-1}}$ decreases rapidly with every bit-flip that separates $\ell_i$ from $\ell_{i-1}$. Indeed, $\log(10^{-2})$ lies below the threshold identified in Eq.~\eqref{eq:threshold}, so if two $L^{\ell_i}_{\ell_{i-1}}$ terms differ by this amount then only the larger of the two will be significant.

Using Eq.~\eqref{eq:transition_log_expansion} and the fact that $L^{\ell_i}_{\ell_{i-1}} \equiv G^{\ell_i}_{\ell_{i-1}} + \log J_{\ell_{i-1}\ell_i} + \log \hat{P}(\ell_{i-1})$, it is possible to explain the broad patterns observed in Figure~\ref{fig:contributions}. Beginning with the contributions to $\ket{000}$, the magnitude of $L^0_0$ (blue line in plot \hyperref[fig:contributions]{5A}) will dominate, since the system begins in $\ket{000}$ (large prior) and has a high probability of remaining in $\ket{000}$ (large transition element). All other $L^0_{i-1}$ will involve at least one bit-flip and therefore will not be significant relative to $L^0_0$. By contrast, among the contributions to $\ket{001}$ (plot \hyperref[fig:contributions]{5B}), both $L^1_0$ and $L^1_1$ (blue and orange lines respectively) are significant, as each term has one large component (prior for $L^1_0$ and transition element for $L^1_1$) and one small component (transition element for $L^1_0$ and prior for $L^1_1$). After the bit-flip occurs the measurement likelihoods begin to favor $\ket{001}$ over $\ket{000}$, so the prior terms all shift toward $\ket{001}$ as well.

\subsection{Single-term and two-term filters}\label{sec:term_strategies}

The combined results of Sec.~\ref{sec:log_prob} and Sec.~\ref{sec:logsumexp_approximation} offer a straightforward recipe for implementing the log-probability filter of Eq.~\eqref{eq:first_log}. First, the measurement values are passed through the quadratic expressions in Eqs.~(\ref{eq:log_000}--\ref{eq:log_001}) to compute the measurement log-densities $\log P(M^{(1)}_1M^{(2)}|\ell_i\ell_{i-1})$, which are then added to the transition log-probabilities $\log J_{\ell_{i-1}\ell_i}$ and log-priors $\log \hat{P}(\ell_{i-1})$ to generate the $L^{\ell_i}_{\ell_{i-1}}$ terms. These terms must then be fed into Eq.~\eqref{eq:max_logsumexp}, but we are free to choose how many of them to include based on their relative magnitudes. It is here that the algorithm splits into two different paths depending on whether we favor accuracy or simplicity.

If we desire a filter that is highly accurate, then we can take the two largest values of $L^{\ell_i}_{\ell_{i-1}}$ for each $\ell_i$ and substitute them into Eq.~\eqref{eq:max_logsumexp}. In this scenario the LogSumExp reduces to
\begin{equation}\label{eq:two_term_logsumexp}
    \log \hat{P}(\ell_i) = \log \sum_\text{pair} \exp[L^{\ell_i}_{\ell_{i-1}}] =  L^{\ell_i}_* + \log(1 + \exp[\Delta L]),
\end{equation}
where $\Delta L$ is the (negative) difference between the largest and second-largest $L^{\ell_i}_{\ell_{i-1}}$ terms. While Eq.~\eqref{eq:two_term_logsumexp} can be inconvenient to evaluate exactly, methods have been developed in the context of logarithmic number systems that utilize lookup tables and various interpolation schemes to yield efficient and accurate estimates \cite{Haselman_Beauchamp_Wood_Hauck_Underwood_Hemmert_2005}. By keeping the two largest values of $L^{\ell_i}_{\ell_{i-1}}$ we guarantee that the log-probabilities of the true state and its three nearest states are all correctly propagated into the next time step. The log-probabilities of the other four states will not be computed as accurately, although this is acceptable since these states are not involved in the single-flip transitions which the three-qubit code is designed to detect.

Although the logarithm term in Eq.~\eqref{eq:two_term_logsumexp} can be evaluated in a reasonable manner, it still represents an extra computational step that increases the overall complexity of the algorithm. If simplicity is valued over accuracy, then we can alternatively choose to keep only the largest of the $L^{\ell_i}_{\ell_{i-1}}$ after each time step. In this extreme case the LogSumExp collapses to
\begin{equation}\label{eq:one_term_logsumexp}
    \log \hat{P}(\ell_i) = \log \exp[L^{\ell_i}_*] =  L^{\ell_i}_*,
\end{equation}
and $L^{\ell_i}_*$ simply becomes the new posterior. Figure~\ref{fig:contributions} shows that when $\ell_i$ equals the true state only one $L^{\ell_i}_{\ell_{i-1}}$ will be significant, so nothing important is being ignored there. For states adjacent to the true state, however, we will necessarily be ignoring one of the two significant contributions. While this single-term simplification appears quite restrictive, we show in Sec.~\ref{sec:numerical} that Eq.~\eqref{eq:one_term_logsumexp} achieves high accuracy, though as expected it performs somewhat worse than the two-term approach.

\begin{figure}
    \centering
    \includegraphics[width=\textwidth]{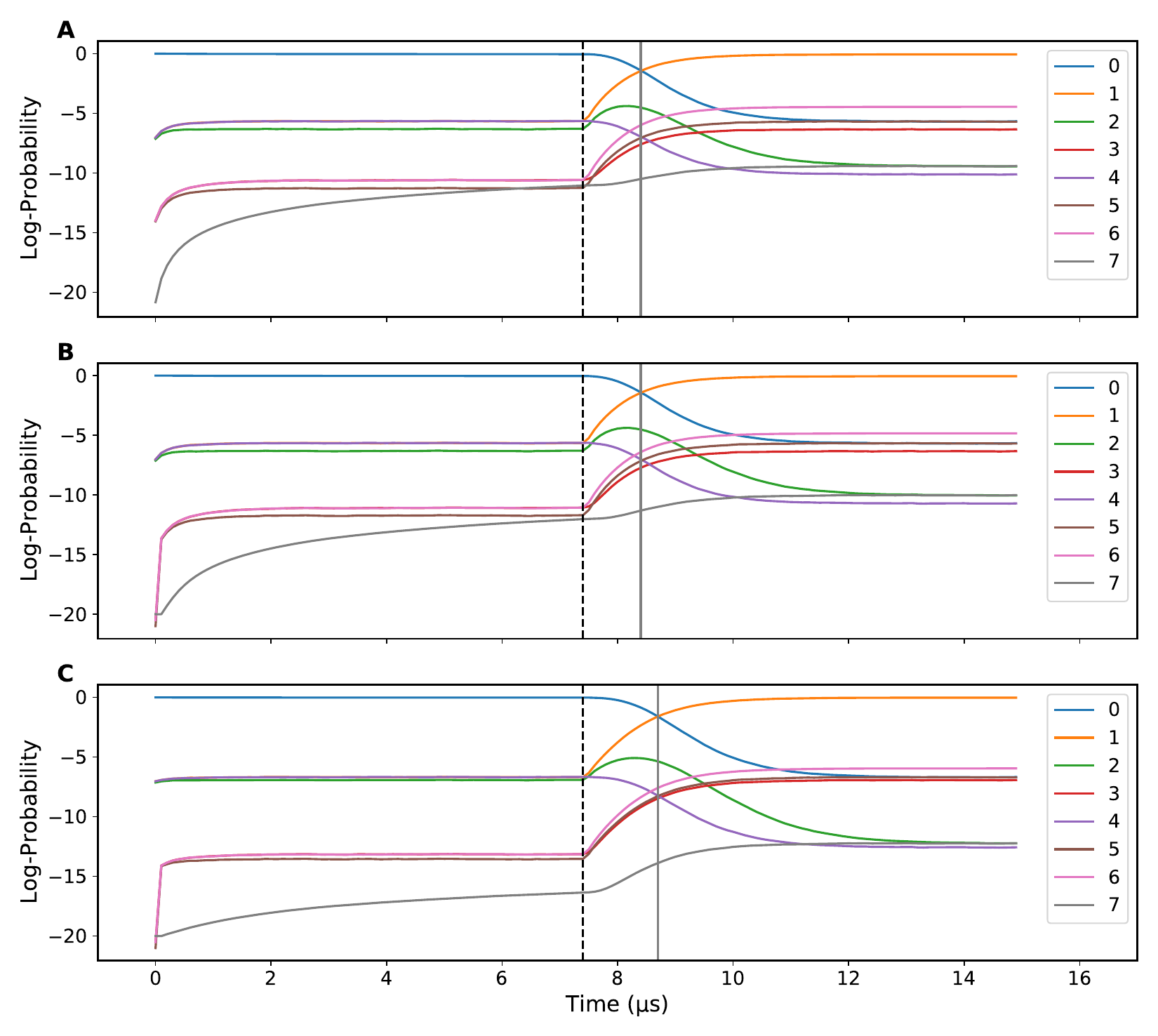}
\caption{Plots of average posterior log-probabilities (normalized) for the optimal filter (plot A, Eq.~\eqref{eq:first_log}), two-term filter (plot B, Eq.~\eqref{eq:two_term_logsumexp}), and single-term filter (plot C, Eq.~\eqref{eq:one_term_logsumexp}), generated from the same parameters and measurement set used in Figure~\ref{fig:contributions}. The solid gray line indicates the time at which the filter correctly identifies that an error has occurred on the third qubit (dashed black line marks the time of the error). The optimal and two-term filter show almost identical behavior, while the single-term filter takes longer to detect the error.}
    \label{fig:log_probs}
\end{figure}

In Figure~\ref{fig:log_probs} we plot the average evolution of the posterior log-probabilities in the presence of an error using outputs from the optimal filter versus those of the single-term and two-term filters. The plots were generated from the same data used to create Figure~\ref{fig:contributions}, and clearly show that all three filters are able to detect the error on the third qubit. The single-term filter performs worst as expected, taking longer to identify the error. Impressively, the two-term filter behaves almost identically to the optimal filter, with small discrepancies emerging for $\ell_i \in \{3, 5, 6, 7\}$ since these states all have more than two significant LogSumExp terms.

\subsection{Normalization}

\begin{figure}
    \centering
    \includegraphics[width=0.9\textwidth]{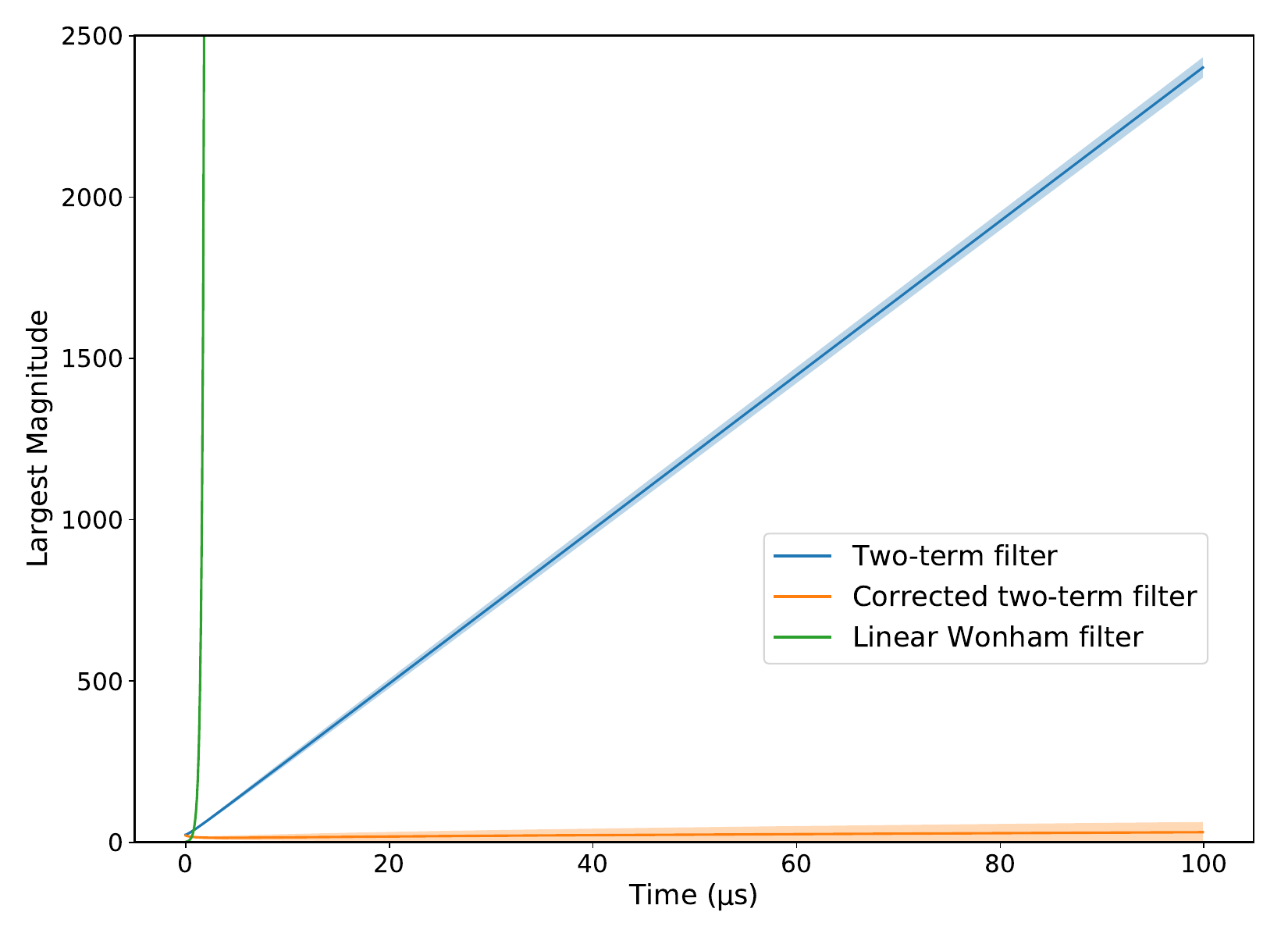}
    \caption{This plot tracks the magnitude of the largest probability output as a function of time for the linear Wonham filter and two different versions of our two-term filter (the single-term filter behaves identically), with $T = 0.1\ \mu \text{s}$, $k = 0.4\ \mu \text{s}$, and $\mu = 0$. The solid lines are averages across $10^5$ trials, while the shaded regions enclose one standard deviation (omitted for the Wonham filter). The blue line records the magnitude of the two-term filter without any normalization, demonstrating a clear linear pattern. The orange line shows the same algorithm with the analytic correction factor from Eq.~\eqref{eq:correction_factor} included, which dramatically reduces the largest magnitude (mean and standard deviation of about 30 after 100 $\mu s$).} 
    \label{fig:norms}
\end{figure}

As discussed in Sec.~\ref{sec:obstacles}, one of the practical challenges of using Bayesian methods is the need for periodic normalization of the probability outputs, without which the values will grow or shrink rapidly. This behavior is shown in Figure~\ref{fig:norms}, where the outputs of the linearized Wonham filter from Eq.~\eqref{eq:linear_wonham} grow exponentially with time. Since exponential changes become linear on a logarithmic scale, the outputs of our logarithmic filters instead scale linearly with time. In situations where error tracking needs to be done for only a short duration, the linear growth of the outputs is likely tolerable and can simply be ignored. This represents a significant improvement over the linearized Wonham filter, which must incorporate a costly normalization routine to be viable over virtually any time scale. 

For longer runs, or in cases where the range of output values is limited due to hardware restrictions, we may wish to slow the growth of the output magnitudes even further. This can be achieved by calculating the average rate of change per time step and simply subtracting this quantity at the end of each step. The average change $\Delta$ is given by
\begin{equation}\label{eq:correction_factor}
    \Delta = -[1 + \log(2\pi\frac{k}{T}) - 3\log \cosh(\mu T) + 3\mu T],
\end{equation}
which is the sum of $\log J_{\ell\ell}$ and $G^\ell_\ell$ when the true state of the system is $\ell$ (since this is the only significant contribution). The corrected outputs are represented by the orange line in Figure~\ref{fig:norms}. Applying this correction greatly slows the growth of the unnormalized log-probabilities, which allows the filters to be run within a very narrow numerical range.

\section{Performance}\label{sec:numerical}

To evaluate the performance of our single-term and two-term filters in an error-correction setting, we simulate a large number of trajectories (see Appendix~\ref{sec_app:simulation}) and then record how accurately the filters are able to correctly identify the final state. For our definition of ``accuracy'' we adopt a binary measure which is 1 when the filter predicts either the true state or a state that differs from the true state by a single bit-flip, and is 0 otherwise. At any given time there will be four states considered to be correct and another four states considered to be incorrect, which gives an expected accuracy of 50\% when guessing randomly. We choose this measure of accuracy because errors on a single qubit can be corrected via simple majority vote, while errors on two or more qubits will signify a logical error which the filter was supposed to have prevented. 

In the following subsections, we present and discuss the performance of the filters under different run durations (Sec.~\ref{sec:time_plots}), error rates $\mu$ (Sec.~\ref{sec:error_plots}), and time steps $T$ (Sec.~\ref{sec:step_plots}). In all simulations we set $k=0.4\ \mu\text{s}$, and the state was initialized to $\ket{000}$. For reference, the performances of our single-term and two-term filters are plotted alongside those of the optimal filter from Eq.~\eqref{eq:bayes_filter} to assess the practical impact of the simplifying assumptions made in Sec.~\ref{sec:term_strategies}. We also include plots for the linearized Wonham filter from Eq.~\eqref{eq:linear_wonham} and the double-threshold scheme from Atalaya and Zhang et al.~\cite{Atalaya_Zhang_Niu_Babakhani_Chan_Epstein_Whaley_2021} to see how well our filters perform relative to existing algorithms. 

\subsection{Experiment duration}\label{sec:time_plots}

\begin{figure}
    \centering
    \includegraphics[width=\textwidth]{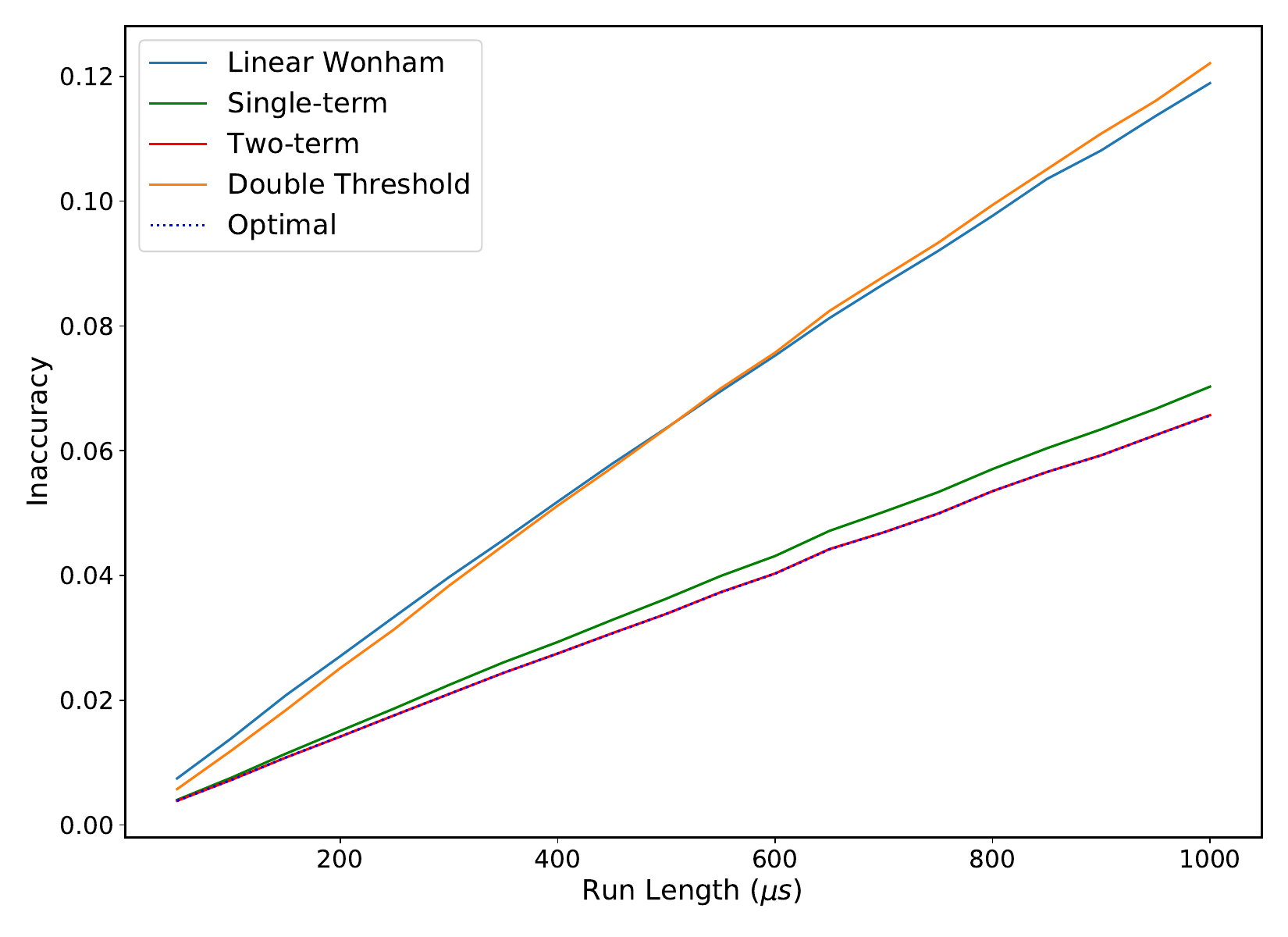}
    \caption{Plot of average inaccuracy (1 - accuracy) of the five different filters as a function of run time, with $k = 0.4\ \mu \text{s}$ and an initial state of $\ket{000}$. The error rate was fixed at $\mu = 2.5\times 10^{-3}\ (\mu \text{s})^{-1}$ and the time step at $T = 10^{-1}\ \mu \text{s}$. The accuracy of the two-term filter is effectively equal to that of the optimal filter, such that the two curves completely overlap, while the single-term version performs slightly worse. The linear Wonham filter and double threshold performed comparably, though both did significantly worse than the logarithmic filters.}
    \label{fig:time_acc}
\end{figure}

Figure~\ref{fig:time_acc} shows the performance of our single-term and two-term filters alongside the reference algorithms as a function of run duration, with the longest runs lasting a full millisecond. The accuracy is a decreasing function of time, since the probability of a logical error will increase as the run grows longer. Mohseninia et al.~\cite{Mohseninia_Yang_Siddiqi_Jordan_Dressel_2020} showed that the accuracy will decrease linearly with run duration at small error rates, a pattern that is clearly visible in our data across all filters.

As expected, the optimal filter achieves the highest accuracy, though it is almost exactly matched by the two-term filter. Despite its greater simplicity, the single-term filter performs within a percentage point of the optimal model. By contrast, the linear Wonham filter and double threshold algorithm both perform significantly worse, though for different reasons. The linearized Wonham filter is constructed explicitly from a first-order approximation, which is exact as $T \rightarrow 0$ but suboptimal for any finite step size. The double threshold scheme, by contrast, is simply suboptimal by design, as it does not directly incorporate the error probabilities or the Gaussian distribution of the noise into its filter.

\subsection{Error rate}\label{sec:error_plots}

\begin{figure}
    \centering
    \includegraphics[width=\textwidth]{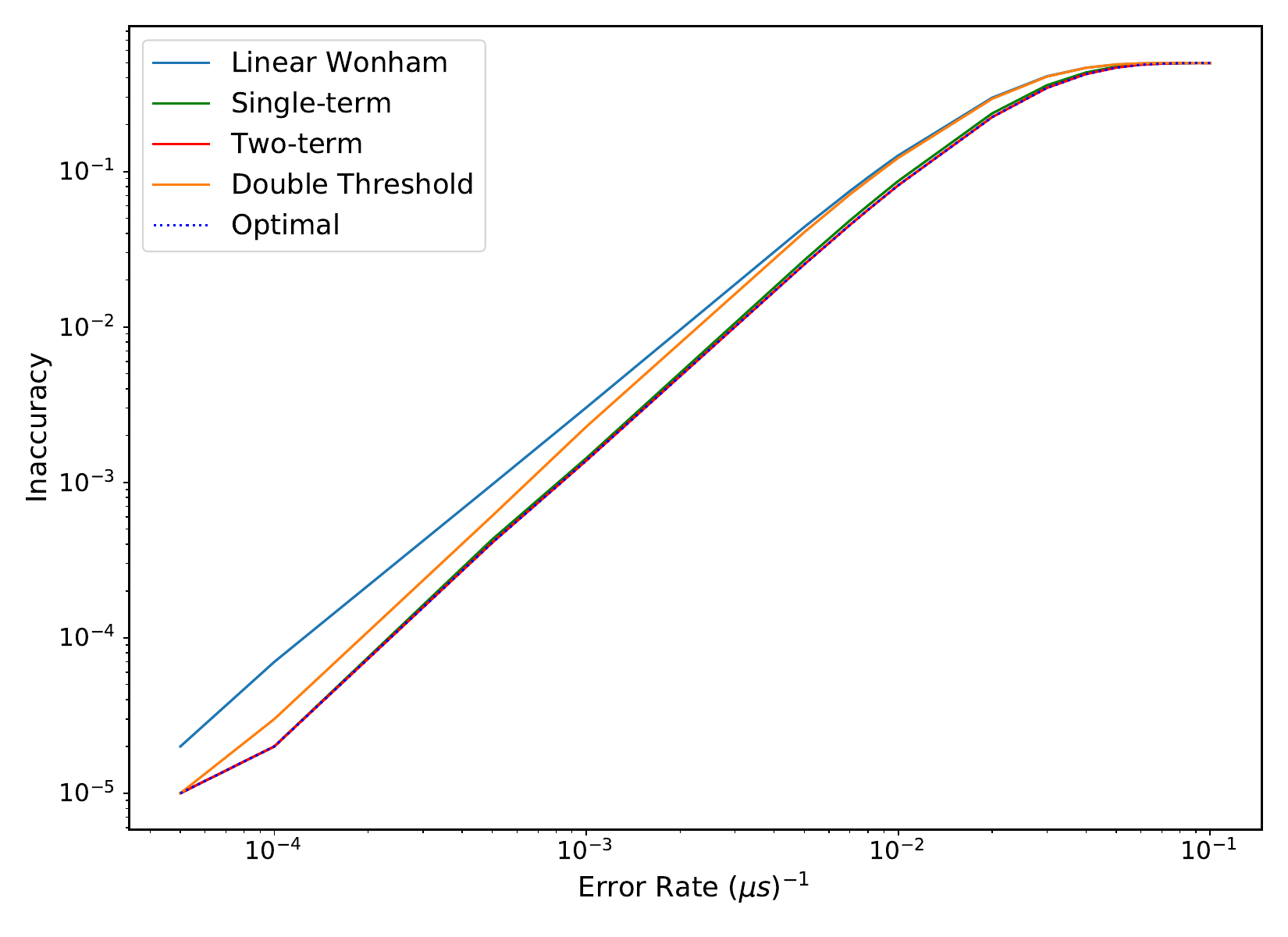}
    \caption{Plot of average inaccuracy (1 - accuracy) of the five filters as a function of error rate, with both axes on a log scale. The duration of the runs was fixed at 100 $\mu \text{s}$ and the time step at $T = 10^{-1}\ \mu\text{s}$, with $k = 0.4\ \mu \text{s}$ and an initial state of $\ket{000}$. The accuracy of the two-term filter is effectively equal to that of the optimal filter (curves overlap completely), with the single-term version performing slightly worse. The linear Wonham filter performed worse than the double threshold, with the gap widening at low error rates.}
    \label{fig:error_acc}
\end{figure}

Figure~\ref{fig:error_acc} shows the accuracy of the five filters as a function of the per-qubit bit-flip error rate $\mu$. On a log-log plot the dependence between the accuracy and error rate is linear for small error rates, but levels off at 0.5 as the error rate approaches $0.1\ (\mu s)^{-1}$. In this high-error regime the system experiences enough logical errors that all memory of the initial state is lost, and the system is therefore unable to distinguish the true state from its complement. As before, the optimal and two-term filters perform almost identically, while the single-term term algorithm is slightly less accurate. The negative impact of non-zero $T$ on the linear Wonham filter is especially obvious at small error rates, where the accuracies of the other filters all converge together while the Wonham filter performs significantly worse. 

\subsection{Time step}\label{sec:step_plots}

\begin{figure}
    \centering
    \includegraphics[width=\textwidth]{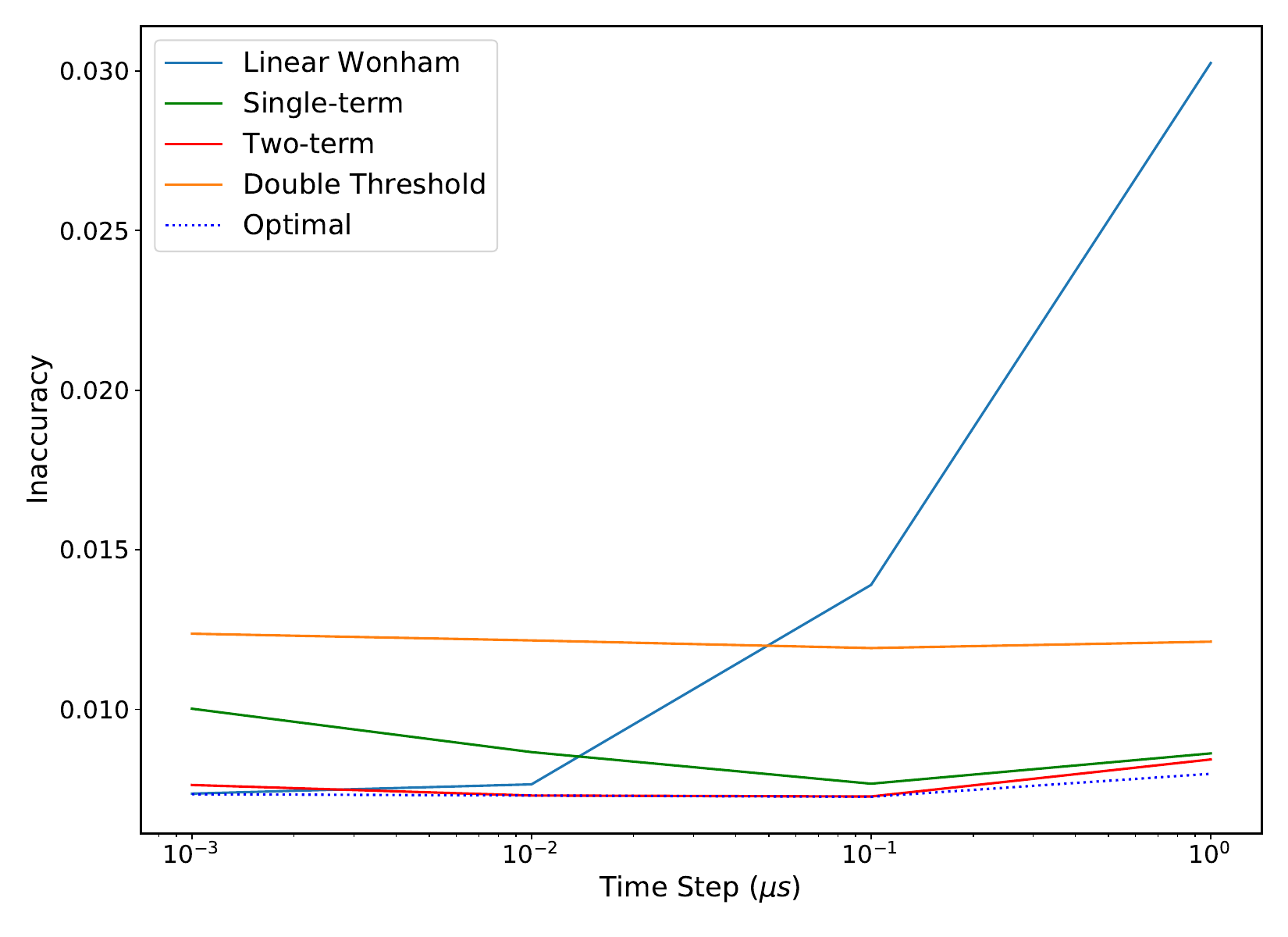}
    \caption{Plot of average inaccuracy (1 - accuracy) of the five different filters as a function of time step $T$, with $k = 0.4\ \mu \text{s}$ and an initial state of $\ket{000}$. The duration of the run was fixed at 100 $\mu \text{s}$ and the error rate at $\mu = 2.5\times 10^{-3}\ (\mu \text{s})^{-1}$. The accuracy of the two-term filter matches that of the optimal curve at most step sizes, but performs slightly worse at very small and very large values of $T$. The single-term filter shows a significant improvement as the time step increases, while the accuracy of the linear Wonham filter decreases rapidly. The double threshold appears largely unaffected by the step size.}
    \label{fig:step_acc}
\end{figure}

Lastly, Figure~\ref{fig:step_acc} shows the accuracy of the filters as a function of the measurement integration time $T$, which reveals many different trends. The time steps range from $1$~ns to 1~$\mu s$, mapping to noise variances on the order of $100$ to $0.1$ respectively, which allows us to explore the behavior of these filters under both high-noise and low-noise conditions. 

Focusing first on the double-threshold scheme, it is clear that the step size does not have a significant effect on its performance. This is unsurprising, as the thresholding procedure of \cite{Atalaya_Zhang_Niu_Babakhani_Chan_Epstein_Whaley_2021} already includes an exponential moving average which smooths over the Gaussian noise for all step sizes. Indeed, since smaller values of $T$ result in more measurements per time interval, the stronger noise is balanced out by the greater sample size of the average. This is the same trade-off described in Eq.~\eqref{eq:standard_error} in the context of mean value estimation.

Looking next at the linear Wonham filter, Figure~\ref{fig:step_acc} provides a very clear demonstration of its dependence on $T$. At a time step of $1$ ns the filter is effectively optimal, which is expected given that the differential Wonham filter of Eq.~\eqref{eq:wonham_filter} is known to be exact. However, under more realistic conditions in which hardware latency and other factors generate longer time steps, the first-order nature of the filter starts to negatively impact its performance.

For the single-term and two-term filters, several different trends emerge. The single-term filter (green line) has a much poorer performance at small time steps than the two-term filter (red line), while at larger time steps their performances converge. This occurs because the transition probabilities are suppressed as $T \rightarrow 0$, which means that almost all of the contributions to the LogSumExp are small. For the two-term filter this is not a significant problem, since taking both the largest and second-largest terms allows for the log-probability to accumulate over time. The outputs of the single-term filter cannot build up in this manner, so the log-probabilities are reduced far below their true values. As $T$ increases this effect is diminished, so the performance of the two filters will begin to converge. Figure~\ref{fig:step_acc} also shows that the two-term filter diverges from the optimal filter at large $T$, since the single-error Gaussian approximation made in Sec.~\ref{sec:single_error} starts to break down when $\mu T$ is large and $\frac{k}{T}$ is small.

\section{Discussion}\label{sec:discussion}

The core objective of our work here was to devise a quantum error detection filter that is grounded in formal Bayesian analysis, yet also practical to implement on real-world quantum hardware. We have assumed that the continuous measurements are represented as a set of discrete signal averages taken over a time $T$, which then serve as inputs to the filter. This finite time step $T$ cannot be avoided when working with actual hardware, since even if weak measurements could be made on arbitrarily short time scales there would still be latency associated with transferring and processing the data. As discussed in Sec.~\ref{sec:measurement}, the fact that $T$ must be non-zero has important effects on the theoretical distribution of the measurement values, which become more significant as both $T$ and the error rate $\mu$ increase.

The filter derived from our Bayesian analysis in Sec.~\ref{sec:optimal} can be understood as as a discretized version of the Wonham filter for finite $T$, though our derivation did not originate from its stochastic differential equation. Much of the difficulty in that analysis stemmed from the fact that errors can occur within the integration period of the measurement, which causes the underlying syndrome means to be distributed across the entire $[-1, 1]$ interval instead of being constrained to only $\pm 1$. Even with this additional degree of freedom, a closed-form solution for the marginal distribution of either syndrome can be derived by simply summing Eq.~\eqref{eq:single_error_1} over $0 \leq e_k < \infty$. The truly challenging part of the analysis comes when the two syndrome distributions are coupled together by errors on the second qubit, which induces complex inter-dependencies between $\bar{S}^{(1)}_i$ and $\bar{S}^{(2)}_i$. One avenue for future work could lie in analyzing the form of these joint distributions, with the goal of determining whether a convenient or illuminating analytic form exists. In the absence of such a form, some type of simplification will always be necessary in order to utilize the filter.

Indeed, the single-error assumption that we made when deriving our log-probability filters in Sec.~\ref{sec:practical_filter} was designed explicitly to avoid the mathematical challenges associated with the joint syndrome distributions. This assumption was the most restrictive that we made when developing the filters, and it becomes increasingly inaccurate as the average number of errors per step ($\mu T)$ increases. While it is true that the error tracking problem as a whole becomes quite challenging when $\mu T$ is large, the question of how a filter could best be designed for such a situation warrants further exploration.

However, in the preferred regime where $\mu T$ is small, the numerical results in Sec.~\ref{sec:numerical} show that our single-term and two-term log-probability filters are highly effective. In every test that we conducted, these two filters outperformed both the first-order Wonham filter and double threshold algorithm by a significant margin, with the two-term filter being virtually optimal over a wide range of $\mu$ and $T$ values. The single-term filter performs slightly worse, but is surprisingly effective given its simplicity. Furthermore, Figure~\ref{fig:step_acc} shows that the performance of the single-term filter increases as $T$ grows larger, up to a value of about $10^{-1}\ \mu\text{s}$. This, combined with the fact that the optimal filter experiences only a negligible improvement in performance when moving to smaller $T$, suggests that effort spent on latency reduction is likely to provide diminishing returns for filter accuracy. 

Given the rapidly growing size of modern quantum hardware, it is safe to assume that error correction will remain an integral component of quantum computation for the foreseeable future. The three-qubit toy model analyzed here is insufficient to protect against arbitrary errors, so an obvious extension of our work would focus on developing continuous error correction filters for larger codes e.g., the Shor, Steane, or subsystem codes which provide more comprehensive protection \cite{Shor_1995}\cite{Steane_1996}. Our results suggest that Bayesian filters similar to those presented here would find great success on these more complex systems.

\section*{Acknowledgements}

I.C. was supported by the National Aeronautics and Space Administration under Grant/Contract/
Agreement No. 80NSSC19K1123 issued through the Aeronautics Research Mission Directorate.
This material is also based upon work supported by the U.S. Department of Energy, Office of Science, National Quantum Information Science Research Centers, Quantum Systems Accelerator.

\printbibliography

\appendix
\section{Appendix}\label{sec:appendix}

\subsection{Syndrome distributions}\label{sec_app:syndrome_dist}

\subsubsection{General form}\label{sec_app:synd_general}

In Eq.~\eqref{eq:synd_factorize}, we expand the syndrome distribution $P(\bar{S}^{(1)}_i\bar{S}^{(2)}_i|\ell_i\ell_{i-1})$ in terms of the error probabilities $p(e_k|\ell_i\ell_{i-1})$ and the conditional distribution $P(\bar{S}^{(1)}_i\bar{S}^{(2)}_i|\ell_{i-1};e_1e_2e_3)$ of observing a particular pair of syndrome means given a number of errors $e_k$ on the $k$th qubit. Figure~\hyperref[fig:syndromes]{1C} shows that the values of $\bar{S}^{(1)}_i$ and $\bar{S}^{(2)}_i$ depend explicitly on the number and \textit{location} of the bit-flip errors within the integration interval, so we introduce a new set of variables $\{x^{k_j}_j\}$ which denote the positions of the $N \equiv e_1 + e_2 + e_3$ errors. The index $j$ indicates the order that the errors occur, such that $x^{k_{j-1}}_{j-1} \leq x^{k_j}_j$, while $k_j \in \{1, 2, 3\}$ denotes on which qubit the error occurs. Note that these ``positions'' are in fact times $t$ during the run, but for convenience we define the beginning of the $i$th integration period to be at $t = 0$ and then divide the times by $T$ so that they all lie on the unit-less interval $[0, 1]$. The value $0 \leq x^{k_j}_j \leq 1$ then denotes a location within this interval. Using this new set of variables, the syndrome density is given by the marginalization over all error locations and all possible qubit assignments:
\begin{equation}\label{eq_app:synd_dist_integral}
\begin{split}
    P(\bar{S}^{(1)}_i&\bar{S}^{(2)}_i|\ell_{i-1};e_1e_2e_3) = 
    \\
    &\sum^{3}_{k_1 = 1}...\sum^{3}_{k_N = 1}\int^1_0dx^{k_1}_1\int^{1}_{x^{k_1}_1}dx^{k_2}_2...\int^1_{x^{k_{N-1}}_{N-1}}dx^{k_N}_NP(\bar{S}^{(1)}_i\bar{S}^{(2)}_i|\ell_{i-1}\{x^{k_j}_j\})P(\{x^{k_j}_j\}|e_1e_2e_3),
\end{split}
\end{equation}
where we have now further factorized the syndrome density into the product of two new distributions. 

Starting first with $P(\{x^{k_j}_j\}|e_1e_2e_3)$, we know that the location of an error and the qubit that it is assigned to are independent, so we have
\begin{equation}\label{eq_app:error_loc_factors}
    P(\{x^{k_j}_j\}|e_1e_2e_3) = P(\{x_j\}|e_1e_2e_3)P(\{k_j\}|e_1e_2e_3).
\end{equation}
Given that each qubit is equally likely to experience an error, $P(\{k_j\}|e_1e_2e_3)$ must be uniform over all \textit{valid} assignments of $\{k_j\}$, where an assigment is valid if exactly $e_k$ errors are assigned to the $k$th qubit. Since the distribution must be normalized, each configuration will have a probability equal to one over the total number of valid configurations:
\begin{equation}\label{eq_app:error_qubit_dist}
    P(\{k_j\}|e_1e_2e_3) = 
    \begin{cases}
        \dfrac{e_1!e_2!e_3!}{N!} & \text{when valid}
        \\
        0 & \text{otherwise}.
    \end{cases}
\end{equation}
Similarly, the errors themselves will be uniformly distributed within the interval, so the joint distribution of all error locations $\{x_j\}$ is one over the integral across all positions
\begin{equation}\label{eq_app:error_loc_dist}
    P(\{x_j\}|e_1e_2e_3)  = \left[ \int^1_0dx^{k_1}_1\int^{1}_{x^{k_1}_1}dx^{k_2}_2...\int^1_{x^{k_{N-1}}_{N-1}}dx^{k_N}_N \right]^{-1} = N!,
\end{equation}
which depends only on the total number of errors.

Turning now to $P(\bar{S}^{(1)}_i\bar{S}^{(2)}_i|\ell_{i-1}\{x^{k_j}_j\})$, the location of the errors and their qubit assignments will completely determine the value of the syndromes, so we will have a product of Dirac delta distributions. Specifically, the average value over an interval depends on the gaps in time between successive errors, as well as the gap between the start of the interval and the first error and the gap between the end of the interval and the last error. As shown in Figure~\hyperref[fig:syndromes]{1C}, the syndrome value is given by the sum of these gaps, with the sign of each contribution alternating due to the parity flips caused by the errors. Therefore, the syndrome distribution has the form
\begin{equation}\label{eq_app:synd_delta}
\begin{split}
    P(\bar{S}^{(m)}_i|\ell_{i-1}\{x^{(m)}_j\}) &= \delta\left(\pm_m\bar{S}^{(m)}_i - x^{(m)}_1 - \left[\sum^{N^{(m)}}_{j = 2}(-1)^{j-1}(x^{(m)}_j - x^{(m)}_{j-1})\right] - (-1)^{N^{(m)}}(1 - x^{(k)}_{N^{(m)}})\right)
    \\
    &= \delta\left(\pm_m\bar{S}^{(m)}_i - (-1)^{N^{(m)}} - 2\sum^{N^{(m)}}_{j = 1} (-1)^{j-1}x^{(m)}_j\right),
\end{split}
\end{equation}
where $\{x^{(m)}_j\}$ is the set of errors which affect the $m$th parity operator and $N^{(m)}$ is the number of errors in this set. For the second interval in Figure~\hyperref[fig:syndromes]{1C}, the syndrome distributions would then be given by
\begin{equation}
    P(\bar{S}^{(m)}_i|\ell_{i-1}\{x^{(m)}_j\}) = \delta\left(\pm_m\bar{S}^{(m)}_i - x_1 + (1 - x_1) \right) = \delta\left(\pm_m\bar{S}^{(m)}_i - 0.4\right),
\end{equation}
where $x_1 = 0.7$ is the normalized position of the error on the second qubit.

Using Eqs.~(\ref{eq_app:error_loc_factors}--\ref{eq_app:synd_delta}), we can rewrite Eq.~\eqref{eq_app:synd_dist_integral} as 
\begin{equation}\label{eq_app:syndrome_dist_integral_final}
\begin{split}
    P(&\bar{S}^{(1)}_i\bar{S}^{(2)}_i|\ell_{i-1};e_1e_2e_3) = 
    \\
    &e_1!e_2!e_3!\sum_{\{k_j\}\text{ valid }}\int^1_0dx^{k_1}_1\int^{1}_{x^{k_1}_1}dx^{k_2}_2...\int^1_{x^{k_{N-1}}_{N-1}}dx^{k_N}_N P(\bar{S}^{(1)}_i|\ell_{i-1}\{x^{(1)}_j\})P(\bar{S}^{(2)}_i|\ell_{i-1}\{x^{(2)}_j\}),
\end{split}
\end{equation}
which can be understood as the sum of volume integrals over the values of $\{x^{(k_j)}_j\}$ which satisfy the Dirac distributions from Eq.~\eqref{eq_app:synd_delta}. In principle, all that remains is to solve for the appropriate integral boundaries and then evaluate the integrals, though this is difficult to do for an arbitrary distribution of errors. In Sec.~\ref{sec_app:single_syndrome} we describe how Eq.~\eqref{eq_app:syndrome_dist_integral_final} can be evaluated when errors occur on only a single qubit, while in Sec.~\ref{sec_app:multi-qubit} we discuss the challenges of generalizing this to errors on multiple qubits, specifically when the second qubit is involved.

\subsubsection{Distributions for single-qubit errors}\label{sec_app:single_syndrome}

If errors are constrained to occur on only a single qubit, then one of the syndrome means will behave in a trivial manner. For errors on the first qubit we will have $\bar{S}^{(2)}_i = \pm_2 1$, and for errors on the the third qubit we will have $\bar{S}^{(1)}_i = \pm_1 1$. With errors on the second qubit both syndromes will change in an identical manner, so we can choose to model explicitly the behavior of $\bar{S}^{(1)}_i$ and then just set $\bar{S}^{(2)}_i = \bar{S}^{(1)}_i$.

We begin by considering a scenario where the first qubit experiences $N$ errors, while $e_2 = e_3 = 0$. Under these conditions, $P(\bar{S}^{(2)}_i|\ell_{i-1}\{x^{(2)}_j\}) = \delta (\bar{S}^{(2)}_i \mp_2 1)$, as $\{x^{(2)}_j\}$ is empty and thus there are no errors on the second or third qubits to flip the initial syndrome of the second parity operator. Indeed, the syndrome distribution of Eq.~\eqref{eq_app:syndrome_dist_integral_final} factorizes completely, with the marginal distribution of $\bar{S}^{(1)}_i$ given by
\begin{equation}\label{eq_app:synd_1_integral}
    P(\bar{S}^{(1)}_i|\ell_{i-1};N00) = 
    N!\int^1_0dx_1\int^{1}_{x_1}dx_2...\int^1_{x_{N-1}}dx_N \delta\left(\pm_1\bar{S}^{(1)}_i - (-1)^N - 2\sum^N_{j = 1} (-1)^{j-1}x_j\right).
\end{equation}
Note that the sum of Eq.~\eqref{eq_app:syndrome_dist_integral_final} has collapsed into a single term, since there is only one way to assign $N$ errors to a single qubit. Given that $k_j = 1$ for all $j$, we have removed the superscripts from the error locations.

As mentioned previously, we can view Eq.~\eqref{eq_app:synd_1_integral} as a volume integral over the regions of the variable space which give the desired value of $\bar{S}^{(1)}_i$. To derive the integration bounds, we consider first the constraints placed on $x_1$. The first error will contribute $\pm_1 (x_1 - 0) = \pm_1 x_1$ to the value of the syndrome, which cannot be altered by any subsequent errors. Therefore, the range of $\bar{S}^{(1)}_i$ after the first error is
\begin{equation}
    \pm_1x_1 - (1 - x_1) \leq \bar{S}^{(1)}_i \leq \pm_1x_1 + (1 - x_1),
\end{equation}
since the rest of interval can be entirely negative or entirely positive at the two extremes. Writing the limits with respect to $x_1$ gives
\begin{equation}
    x_1 \leq \frac{1 \pm_1 \bar{S}^{(1)}_i}{2},
\end{equation}
which becomes the upper integration bound of the first integral. We can repeat this same process for each $x_j$, taking the sum of the contributions from errors up to $x_j$ and either adding or subtracting the remaining interval length to get the upper and lower bounds respectively on $\bar{S}^{(1)}_i$. These bounds are then rearranged to derive an upper bound on $x_j$, which has the generic form
\begin{equation}\label{eq_app:bound}
    B_j \equiv \frac{1}{2} + (-1)^{j - 1}\left(\frac{\pm_1 \bar{S}^{(1)}_i}{2} + \sum^{j-1}_{n=1}(-1)^nx_n\right),
\end{equation}
such that the integration limits are $x_{j-1} \leq x_j \leq B_j$. These bounds have the recursion relation
\begin{equation}\label{eq_app:bound_recursion}
    B_j = (1 + x_{j-1}) - B_{j-1}
\end{equation}
which is key to evaluating the integrals.

Using Eq.~\eqref{eq_app:bound} and Eq.~\eqref{eq_app:bound_recursion}, the volume integral in Eq.~\eqref{eq_app:synd_1_integral} can be easily solved. The integral over $x_N$ will collapse to $\frac{1}{2}$ due to the Dirac delta function, and contribute nothing further to the volume. Considering the next two integrals we have
\begin{equation}
\begin{split}
    \int^{B_{N-2}}_{x_{N-3}}dx_{N-2}\int^{B_{N-1}}_{x_{N-2}}dx_{N-1} &= \int^{B_{N-2}}_{x_{N-3}}dx_{N-2}(B_{N_1} - x_{N-2}) 
    \\
    &= (1 - B_{N-2})\int^{B_{N-2}}_{x_{N-3}}dx_{N-2}
    \\
    &=(1 - B_{N-2})(1 - B_{N-3}),
\end{split}
\end{equation}
where we used the recursive expression $B_j - x_{j-1} = 1 - B_{j-1}$. The result of the first integral does not have any dependence on $x_{N-2}$ and therefore is unaffected by the second integral. Indeed, the generic action of the integral for $x_{j-1}$ on $(1 - B_j)^a$ is
\begin{equation}\label{eq_app:integral_map}
    \int^{B_{j-1}}_{x_{j-2}}dx_{j-1}(1 - B_j)^a = \frac{1}{a + 1}(1 - B_{j-2})^{a+1},
\end{equation}
which maps $B_j \rightarrow B_{j-2}$. By repeated application of Eq.~\eqref{eq_app:integral_map}, the probability density in  Eq.~\eqref{eq_app:synd_1_integral} reduces to
\begin{equation}\label{eq_app:integral_evaluated}
    P(\bar{S}^{(1)}_i|\ell_{i-1};N00) \propto (1 - B_0)^{N//2}(1 - B_1)^{(N-1)//2},
\end{equation}
where $x//y$ is $x$ floor-divided by $y$, and where $B_0$ is defined from the recursion relation in Eq.~\eqref{eq_app:bound_recursion} with $j = 1$ and $x_0 \equiv 0$. Substituting the expressions for $B_1$ and $B_0$ into Eq.~\eqref{eq_app:integral_evaluated} gives
\begin{equation}
    P(\bar{S}^{(1)}_i|\ell_{i-1};N00) = \frac{N!}{2(N//2)!((N-1)//2)!}(\frac{1 \pm_1 \bar{S}^{(1)}_i}{2})^{N//2}(\frac{1\mp_1 \bar{S}^{(1)}_i}{2})^{(N-1)//2}
\end{equation}
after normalization, which is simply a beta distribution with respect to $\frac{1 \pm \bar{S}^{(1)}_i}{2}$ on the interval $[-1, 1]$. This result can be easily generalized to $N$ errors on the third qubit by replacing $\bar{S}^{(1)}_i$ with $\bar{S}^{(2)}_i$ and $\pm_1$ with $\pm_2$.

\subsubsection{Challenges of multi-qubit errors}\label{sec_app:multi-qubit}

When errors occur on the second qubit and either the first or third qubits (or both), the evaluation of Eq.~\eqref{eq_app:synd_dist_integral} is far more challenging. First, the sum over $\{k_j\}$ will not collapse to a single term, so we must evaluate multiple volume integrals. More importantly, the error locations assigned to the second qubit in each of those volume integrals will have to simultaneously satisfy constraints imposed by both $\bar{S}^{(1)}_i$ and $\bar{S}^{(2)}_i$, which means that the bounds $B_j$ on those errors will be given by
\begin{equation}
    B_j = \min\left[\frac{1}{2} + (-1)^{j - 1}\left(\frac{\pm_1 \bar{S}^{(1)}_i}{2} + \sum^{N^{(1)}_j}_{n=1}(-1)^nx^{(1)}_n\right), \frac{1}{2} + (-1)^{j - 1}\left(\frac{\pm_2 \bar{S}^{(2)}_i}{2} + \sum^{N^{(2)}_j}_{n=1}(-1)^nx^{(2)}_n\right)\right],
\end{equation}
where $N^{(m)}_j$ is the number of errors occurring earlier than $x^{k_j}_j$ that affect the $m$th parity operator. These bounds create dependencies between the variables in $\{x^{(1)}_j\}$ and $\{x^{(2)}_j\}$ which prevent us from using the simple recursion strategy employed in Eq.~\eqref{eq_app:integral_map}. That said, an expression for $P(\bar{S}^{(1)}_i\bar{S}^{(2)}_i|\ell_{i-1};e_1e_2e_3)$ can still be calculated from Eq.~\eqref{eq_app:synd_dist_integral} in a brute-force manner using symbolic computation tools, though this becomes increasingly impractical as the values of $e_1$, $e_2$, and $e_3$ grow.  

\subsection{Implementing the optimal filter}\label{sec_app:implement_filter}

From Eq.~\eqref{eq:bayes_filter}, the optimal Bayesian filter is a function of the measurement density $P(M^{(1)}_iM^{(2)}_i|\ell_i\ell_{i-1})$ and transition elements $J_{\ell_{i-1}\ell_i}$. The matrix $J$ is easy to compute, but in order to calculate $P(M^{(1)}_iM^{(2)}_i|\ell_i\ell_{i-1})$ we must evaluate the integral in Eq.~\eqref{eq:measurement_dist}. Although the analytic solution is not available to us, we can approximate the value of the integral using a Riemann sum
\begin{equation}\label{eq_app:riemann}
\begin{split}
    &P(M^{(1)}_iM^{(2)}_i|\ell_i\ell_{i-1}) \approx
    \\
    &\frac{1}{n^2}\sum^{n-1}_{m^{(1)}=-n}\sum^{n-1}_{m^{(2)}=-n}\prod^2_{j=1}\frac{\exp[\frac{-T}{2k}(M^{(j)}_i - ( \frac{m^{(j)} + \frac{1}{2}}{n}))^2]}{\sqrt{2\pi\frac{k}{T}}} P(\bar{S}^{(1)}_i = \frac{m^{(1)} + \frac{1}{2}}{n}, \bar{S}^{(2)}_i = \frac{m^{(2)} + \frac{1}{2}}{n}|\ell_i\ell_{i-1}),
\end{split}
\end{equation}
where the interval $[-1, 1]$ has been discretized into $4n^2$ evenly-spaced segments. 

The syndrome density is estimated by using Eq.~\eqref{eq:synd_factorize} to factorize it into $P(\bar{S}^{(1)}_i\bar{S}^{(2)}_i|e_1e_2e_3)$ and $P(e_1e_2e_3|\ell_i\ell_{i-1})$, the latter of which can be evaluated analytically. To approximate the value of $P(\bar{S}^{(1)}_i\bar{S}^{(2)}_i|e_1e_2e_3)$, we construct histograms of $\bar{S}^{(1)}_i$ and $\bar{S}^{(2)}_i$ with $4n^2$ bins for values of $e_1$, $e_2$, and $e_3$ such that $e_1 + e_2 + e_3 \leq N$. The value of $N$ is chosen so that the probability of experiencing more than $N$ errors in a given interval is below some threshold value. Since $\mu$ and $T$ only impact $P(e_1e_2e_e|\ell_i\ell_{i-1})$, these histograms can be reused for multiple error rates and integration times. To reconstruct $P(\bar{S}^{(1)}_i\bar{S}^{(2)}_i|\ell_i\ell_{i-1})$, we perform the sum in Eq.~\eqref{eq:synd_factorize} using the histograms and the analytic error probabilities to generate a single combined histogram.

The sum in Eq.~\eqref{eq_app:riemann} must be evaluated at every time step, which becomes a significant computational bottleneck for large $n$. To speed up the filter, we constructed a lookup table for the Gaussian distribution at discrete values across $M^{(1)}_i$ and $M^{(2)}_i$ and then used bi-linear interpolation to compute values for continuous inputs. Given the smoothness of the Gaussian function, these interpolations were highly accurate and very fast to compute.

\subsection{Convergence to the Wonham filter}\label{sec_app:converge}

The (unnormalized) Bayesian filter given in Eq.~\eqref{eq:bayes_filter} can be shown to converge to the linear Wonham filter of Mohseninia et al. \cite{Mohseninia_Yang_Siddiqi_Jordan_Dressel_2020} as $T \rightarrow 0$, and thus to converge to the Wonham filter after normalization. Up to the smallest order in $T$, the transition matrix elements are given by
\begin{equation}\label{eq_app:first_order_J}
    J_{\ell_{i-1}\ell_i} \approx (\mu T)^{d(\ell_{i-1}, \ell_i)}(1 - 3\mu T),
\end{equation}
where $d(\ell_{i-1}, \ell_i)$ is the Hamming distance between the three-bit representations of $\ell_i$ and $\ell_{i-1}$. Ignoring the normalization factor, the measurement density can be expanded up to its smallest order in $T$ as
\begin{equation}\label{eq_app:first_order_meas}
\begin{split}
    P(M^{(1)}_i&M^{(2)}_i|\ell_i\ell_{i-1}) \propto
    \\
    &\int^1_{-1}\int^1_{-1}\left(1 - \frac{T}{2k}\left[(M^{(1)}_i - \bar{S}^{(1)}_i)^2 + (M^{(2)}_i - \bar{S}^{(2)}_i)^2\right]\right)\tilde{P}(\bar{S}^{(1)}_i\bar{S}^{(2)}_i|\ell_i\ell_{i-1})d\bar{S}^{(1)}_id\bar{S}^{(2)}_i,
\end{split}
\end{equation}
where $\tilde{P}(\bar{S}^{(1)}_i\bar{S}^{(2)}_i|\ell_i\ell_{i-1})$ is equal to $P(\bar{S}^{(1)}_i\bar{S}^{(2)}_i)|e_1e_2e_3)$ such that $e_k = 1$ if there is a net flip on the $k$th qubit when moving from $\ell_{i-1}$ to $\ell_i$ and $e_k = 0$ otherwise. This is the syndrome density with the fewest errors that is still consistent with the state transition.

Substituting Eq.~\eqref{eq_app:first_order_J} and Eq.~\eqref{eq_app:first_order_meas} into Eq.~\eqref{eq:bayes_filter} gives
\begin{equation}\label{eq_app:first_order_sum}
    \hat{P}(\ell_i) \propto \sum_{\ell_{i-1}}(\mu T)^{d(\ell_{i-1}, \ell_i)}(1 - 3\mu T)(1 - \frac{T}{2k}I_{\ell_{i-1}\ell_i})P(\ell_{i-1}),
\end{equation}
where $I$ is a matrix with elements given by
\begin{equation}
    I_{\ell_{i-1}\ell_i} = \int^1_{-1}\int^1_{-1}\left[(M^{(1)}_i - \bar{S}^{(1)}_i)^2 + (M^{(2)}_i - \bar{S}^{(2)}_i)^2\right]\tilde{P}(\bar{S}^{(1)}_i\bar{S}^{(2)}_i|\ell_i\ell_{i-1})d\bar{S}^{(1)}_id\bar{S}^{(2)}_i.
\end{equation}
The sum in Eq.~\eqref{eq_app:first_order_sum} can be simplified by keeping only those terms which are at most first-order in $T$. With $\mathcal{I} \equiv \{1, 2, 4\}$ containing the set of indices corresponding to bit-flips on the first, second, and third qubits respectively, we have
\begin{equation}\label{eq_app:first_order_sum_simp}
    \hat{P}(\ell_i) \propto P(\ell_{i-1} = \ell_i) + T\left[\mu \sum_{i \in \mathcal{I}}P(\ell_{i-1} = \ell_i \oplus i) - (3\mu + \frac{1}{2k}I_{\ell_i\ell_i})P(\ell_{i-1} = \ell_i)\right],
\end{equation}
which is equivalent to Eq.~\eqref{eq:linear_wonham} after introducing $Q$ and evaluating $I_{\ell_i\ell_i}$.

\subsection{Gaussian fit of single-error distribution}\label{sec_app:gauss_fit}

In Eq.~\eqref{eq:single_error_deriv} we considered the distribution of $M^{(1)}_i$ when a single error occurs on the first qubit, deriving the expression
\begin{equation}\label{eq_app:single_error_dist}
    P(M^{(1)}_i|\ell_{i-1};100) = \frac{1}{4}\left[\text{erf}(\frac{M^{(1)}_i + 1}{\frac{2k}{T}}) - \text{erf}(\frac{M^{(1)}_i - 1}{\frac{2k}{T}})\right].
\end{equation}
From a practical perspective it would be convenient if the distribution were Gaussian, so we compute the mean value $s$ and variance $\sigma^2$ of Eq.~\eqref{eq_app:single_error_dist} and then construct a Gaussian distribution off of these parameters. This can be done by going back to the integral which generated Eq.~\eqref{eq_app:single_error_dist} and simply switching the order of integration
\begin{align}
    s &= \frac{1}{2}\int^1_{-1}d\bar{S}^{(1)}_i\int^{\infty}_{-\infty}dM^{(1)}_iM^{(1)}_i\frac{\exp[\frac{-T}{2k}(M^{(1)}_i - \Bar{S}^{(1)}_i)^2]}{\sqrt{2\pi\frac{k}{T}}} = \frac{1}{2}\int^1_{-1}\bar{S}^{(1)}_id\bar{S}^{(1)}_i = 0
    \\
    \begin{split}
    \sigma^2 &= \frac{1}{2}\int^1_{-1}d\bar{S}^{(1)}_i\int^{\infty}_{-\infty}dM^{(1)}_i(M^{(1)}_i)^2\frac{\exp[\frac{-T}{2k}(M^{(1)}_i - \Bar{S}^{(1)}_i)^2]}{\sqrt{2\pi\frac{k}{T}}} = \frac{1}{2}\int^1_{-1}[(\bar{S}^{(1)}_i)^2 + \frac{k}{T}]d\bar{S}^{(1)}_i
    \\
    &= \frac{1}{3} + \frac{k}{T}.
    \end{split}
\end{align}

\subsection{Simulating trajectories}\label{sec_app:simulation}

To generate synthetic data of duration $nT$, we must correctly evolve the state of the system across each of the $n$ integration intervals (of length $T$), and then generate an appropriate measurement record. To perform this evolution we utilize the so-called ``jump, no-jump'' approach, where individual bit-flips are sampled at each time step and then applied to the system such that the state remains pure across the run. For particular values of $\mu$ and $T$ the probability of experiencing $e_k$ bit-flip errors on the $k$th qubit is given by the Poisson distribution in Eq.~\eqref{eq:poisson}, which can be easily sampled from using standard mathematical libraries. After $e_1$, $e_2$, and $e_3$ have been determined for each of the $n$ time steps, we apply the errors to our initial state $\ket{000}$ in the proper order to evolve the system.

To generate the corresponding measurements, we first sample the syndrome means $\bar{S}^{(1)}_i$ and $\bar{S}^{(2)}_i$ from the measurement distribution $P(\bar{S}^{(1)}_i\bar{S}^{(2)}_i|e_1e_2e_3)$ conditioned on the number of errors that occurred in the step. This can be done in straightforward manner by sampling $e_k$ uniform values on the interval $[0, 1]$ for each qubit and then sorting them by value in ascending order. To calculate $\bar{S}^{(1)}_i$, the time gaps between errors on the first and second qubit are added and subtracted in an alternating order, with the first gap given by the value of the first error and the last gap given by one minus the value of the last error (see Eq.~\eqref{eq_app:synd_delta} for more detail). After the gaps have been summed the syndrome value is multiplied by $\pm_1$ to take into account the parity of the state at the start of interval. An analogous procedure can be be carried out for $\bar{S}^{(2)}_i$ using $\pm_2$ and the gaps between errors on the second and third qubits. Once the syndrome values have been sampled, the values of $M^{(1)}_i$ and $M^{(2)}_i$ are generated by simply adding Gaussian noise with mean zero and variance $\frac{k}{T}$ to $\bar{S}^{(1)}_i$ and $\bar{S}^{(2)}_i$.

The approach to trajectory simulation described here is more involved than most methods for simulating continuous measurements, in that it treats the syndrome mean as a random variable on the interval [-1, 1] that must be conditionally sampled based on the number of errors that occurred in the interval. This is necessary in order to accurately test the performance of a filter on different values of $T$. Other simulation techniques are usually only concerned with the limiting behavior as $T \rightarrow 0$, and therefore just sample syndrome values of $\pm 1$ based on the state of the system.

\end{document}